\pdfoutput=1
\documentclass[preprint]{JHEP3} 

\JHEPspecialurl{http://jhep.sissa.it/JOURNAL/JHEP3.tar.gz}

\usepackage{epsfig,multicol,bbm}
\usepackage{longtable}
\usepackage{amsthm,latexsym,multibox,amssymb,amsfonts,array}
\usepackage[centertags,intlimits]{amsmath}
\usepackage{eucal}
\usepackage{citesort}
\usepackage{tabularx}
\usepackage{graphicx}
\newcommand\fverb{\setbox\pippobox=\hbox\bgroup\verb}
\newcommand\fverbdo{\egroup\medskip\noindent%
            \fbox{\unhbox\pippobox}\ }
\newcommand\fverbit{\egroup\item[\fbox{\unhbox\pippobox}]}
\newbox\pippobox

\newcommand{\al}{\alpha}

\newcommand{\beq}{\begin{equation}}
\newcommand{\eeq}{\end{equation}}
\newcommand{\beqa}{\begin{eqnarray}}

\newcommand{\eqa}{\end{eqnarray}}
\newcommand{\eqnlab}[1]{\label{eqn:#1}}
\newcommand{\Eqnref}[1]{Eq.~(\ref{eqn:#1})}
\newcommand{\f}{\frac}
\newcommand{\mc}{\mathcal}
\newcommand{\pa}{\partial}
\newcommand{\hs}{\hspace{0.1 cm}}

\newcommand{\be}{\beta}

\newcommand{\nn}{\nonumber}

\newcommand{\mf}{\mathfrak}
\newcommand{\om}{\omega}
\newcommand{\mbb}{\mathbb}

\newcommand{\noi}{\noindent}
\newcommand{\p}{\prime}

\title{
\vskip -50pt
\begin{small}
\hfill DAMTP-2008-40   \\
\hfill ULB-TH/08-13    \\
\vskip  30pt
\end{small}
Coxeter group structure of cosmological billiards on compact spatial manifolds}

\author{Marc H{\footnotesize ENNEAUX}$\!$\footnote{Also at \emph{Centro de Estudios Cient\'{\i}ficos
(CECS), Casilla 1469, Valdivia, Chile}} $\,$ and Daniel P{\footnotesize ERSSON}$\!$\footnote{Also at  \emph{Fundamental Physics, Chalmers University of Technology, SE-412 96, G\"oteborg, Sweden}} \\

Physique Th\'{e}orique et Math\'{e}matique, \\ Universit\'{e} Libre
de Bruxelles \&
 International Solvay Institutes, \\
 ULB-Campus Plaine C.P.231, B-1050 Bruxelles,
Belgium\\

E-mail: \email{henneaux, dpersson@ulb.ac.be}}

\author{Daniel H. W{\footnotesize ESLEY} \\

Centre for Theoretical Cosmology,\\
Department of Applied Mathematics and Theoretical Physics,\\
Cambridge University, Wilberforce Road,\\
Cambridge CB3 OWA, United Kingdom\\

E-mail: \email{D.H.Wesley@damtp.cam.ac.uk}}

\abstract{We present a systematic study of the cosmological billiard structures of Einstein-$p$-form systems in which all spatial directions are compactified on a manifold of nontrivial topology. This is achieved for all maximally oxidised theories associated with split real forms, for all possible compactifications as defined by the de Rham cohomology of the internal manifold. In each case, we study the Coxeter group that controls the dynamics for energy scales below the Planck scale as well as the relevant billiard region. We compare and contrast them with the Weyl group and fundamental domain that emerge from the general BKL analysis. For generic topologies we find a variety of possibilities: (i) The group may or may not be a simplex Coxeter group; (ii) The billiard region may or may not be a fundamental domain. When it is not a fundamental domain, it can be described as a sequence of pairwise adjacent chambers, known as a \emph{gallery}, and the reflections in the billiard walls provide a non-standard presentation of the Coxeter group. We find that it is only when the Coxeter group is a simplex Coxeter group, and the billiard region is a fundamental domain, that there is a correspondence between billiard walls and simple roots of a Kac-Moody algebra, as in the general BKL analysis. For each compactification we also determine whether or not the resulting theory exhibits chaotic dynamics.
}

\keywords{Discrete and Finite Symmetries, Spacetime Singularities}

\begin{document}

\section{Introduction and Motivation}

Generic solutions to Einstein's gravity coupled to dilatons and $p$-forms in the neighbourhood of a spacelike singularity possess a surprisingly rich and complicated structure.  This was first revealed by Belinskii, Khalatnikov and Lifshitz (BKL), who showed that in vacuum four-dimensional Einstein gravity the dynamical approach to the singularity is oscillatory and chaotic \cite{BKL1,BKL2,KhalatnikovReview} (see also \cite{Misner}).  Even though the chaotic oscillations disappear for pure gravity in ten dimensions or higher \cite{Demaret:1986su,Demaret:1986ys}, they are present in the bosonic sectors of all supergravities related to string and M-theory due to the presence of $p$-forms \cite{DamourHenneaux1}. Chaos is revealed by casting the dynamics of gravity at each spatial point as billiard motion in a finite-volume region of an auxiliary hyperbolic space \cite{Chitre,Misner2,DHNReview}. This type of dynamics has been extensively studied in mathematics and is well known to exhibit chaotic behaviour.

\subsection{Cosmological Billiards and Overextended U-Duality}

The BKL analysis reveals a connection between gravitational dynamics and Kac-Moody algebras.  For certain theories, including all maximal supergravities, the region where the billiard dynamics takes place can be identified with a bounded region within the Cartan subalgebra $\mf{h}$ of a Lorentzian Kac-Moody algebra $\mf{g}$. In fact, the billiard motion is confined to the fundamental Weyl chamber in $\mf{h}$ and the geometric reflections of the billiard generate the Weyl group $\mf{W}[\mf{g}]$ of $\mf{g}$. This connects the chaotic nature of certain supergravity theories to the hyperbolic nature of the underlying Kac-Moody algebra \cite{DHNJ}. For example, the algebras whose Weyl groups control the dynamics of the string-related supergravities are $E_{10}$ (SUGRA$_{11}$, Type IIA \& IIB), $BE_{10}$ (Heterotic, Type I) and $DE_{10}$ (pure supergravity), which are all hyperbolic \cite{ArithmeticalChaos}. The Kac-Moody algebras governing the BKL behaviour are closely linked to the ``U-duality'' algebras appearing in compactifications of Einstein-dilaton-$p$-form systems to three dimensions: if $\mf{u}$ is the symmetry algebra in three dimensions, then the Kac-Moody algebra $\mf{g}$ controlling the dynamics in the BKL-limit is the ``overextension'' $\mf{u}^{++}$ of the U-duality algebra $\mf{u}$ \cite{Sophie} (see \cite{HullTownsend,UDualityReview} for details on U-duality)\footnote{We assume the real Lie algebra $\mf{u}$ to be split.  The BKL rules for the general case are given in \cite{MHBJ}.}.
By computing the billiard directly in three dimensions where the U-duality symmetry is manifest, and
using the invariance of the billiard structure under toroidal compactification, it has been shown that taking the BKL-limit precisely mimics the overextension procedure described in \cite{Kac}.\footnote{In \cite{Kac} it is called the ``canonical hyperbolic extension''.}
In this way the BKL-limit ``unveils'' an algebraic structure which could play the role of a fundamental underlying symmetry of the theory \cite{TensionExpansion}. This also lends indirect support for Julia's conjecture that  $E_{10}$ is the symmetry of eleven-dimensional supergravity compactified on $T^{10}$ \cite{Julia,Julia2}.\footnote{In this paper we shall mostly focus on the ``overextended'' Kac-Moody algebras, which are closely linked to compactifications to one (timelike) dimension. For related work on affine Kac-Moody algebras, i.e., extended Lie algebras, in the context of reductions to two dimensions, see, e.g., \cite{Breitenlohner,Nicolai1,Nicolai2,KleinschmidtNicolai,Nassiba,PerryPope1,PerryPope2}. So called ``very extended'' Kac-Moody algebras have also been extensively studied in the literature. Most notably, the Lorentzian Kac-Moody algebra $E_{11}$ has been put forward as a possible underlying symmetry of eleven-dimensional supergravity (and, perhaps, M-theory) \cite{Julia,NicolaiSO(16),West1,Englert1,Englert2}. See also the recent work \cite{Bergshoeff,West2} in favour of this conjecture.}

\subsection{Intermediate Asymptotics}

The original BKL analysis is classical and has been pushed all the way to the singularity \cite{BKL1,BKL2}.  As such it is valid for any spatial topology. In the approach to a spacelike singularity there is an asymptotic decoupling of spatial points, and the dynamics becomes ``ultralocal".  These results (decoupling of spatial points and chaotic oscillations) are by now well supported by extensive analytical and numerical evidence; a non-exhaustive list of references include \cite{Berger:1998vxa,Berger:1998us,Andersson:2000cv,Ringstrom:2000mk,Subcritical,Garfinkle:2003bb,Uggla,Garfinkle,FlagPaper}.

The classical analysis has however obvious limitations and it is not clear what becomes of the BKL results for energy scales above the Planck scale, where quantum gravity effects cannot be ignored. In the standard BKL analysis, which ignores quantum effects, no walls are removed as the big crunch is approached.  But, when some spatial dimensions are compact, there is a wide range of initial conditions for which walls corresponding to massive modes are always subdominant until the universe enters the quantum regime, and these walls are not relevant for the billiard analysis while the universe is described by classical physics \cite{Wesley:2005bd,Wesley:2006cd}. For this broad set of initial conditions, there is no epoch in which the usual classical BKL analysis, with the full set of walls, applies.

For this reason, it is of interest to consider the regime of \emph{intermediate asymptotics} where the curvature is much smaller than the Planck curvature but where the billiard analysis applies (see \cite{Damour:2005zb}).  In that pre-Planck regime, it is not true that the topology of spacetime is irrelevant \cite{Wesley:2005bd}.  A modification might arise in the presence of $p$-forms because the massless spectrum of $p$-forms in the lower-dimensional theory depends on the de Rham cohomology of the internal manifold, and since massless degrees of freedom dominate in the BKL-limit before reaching the Planck scale \cite{Wesley:2005bd}, non-trivial compactification eliminates billiard walls corresponding to degrees of freedom which are rendered massive in the compactification.  Depending on which walls are removed by the compactification, a chaotic theory can be rendered non-chaotic. Note that the suppression of massive modes exhibited in \cite{Wesley:2005bd} is a classical result, which follows from the virial theorem.  However, one could argue that it is in fact also true quantum mechanically (in the intermediate regime where the geometry can be treated as classical but the matter fields are quantum-mechanical), for standard methods reveal that the expected energy density $\rho_{\rm pp}$ of pair-produced particles is at most  \cite{Bir84}
\begin{equation}
\rho_{\rm pp} \sim H^2 e^{-m/H},
\end{equation}
where $H$ is the effective Hubble parameter and $m$ the particle mass.  The exponential suppression of massive mode pair production follows the point particle intuition \cite{Gubser:2003vk}, with an even greater suppression expected for string pair production \cite{Tolley:2005ak}.

There are a number of reasons to better understand the interplay between BKL dynamics and compactification.  On the physics side, the BKL-limit with compact internal spaces is relevant for certain types of cyclic or ``pre-big bang" cosmological models built from string or M-theory (see, e.g., \cite{Khoury:2001wf,Steinhardt:2001st,Buonanno:1998bi,Gasperini:2002bn} and references therein).  Cosmological models with a  big crunch/big bang transition rely on a smooth collapsing phase as they approach the big crunch singularity, hence chaotic BKL oscillations close to the singularity are a potential problem for these models.  If chaos can be removed by interpreting our four-dimensional world as an effective description of a more fundamental higher-dimensional theory where the ``troublesome" billiard walls are eliminated through the cohomology of the internal space, the problems with BKL chaos in these models may be circumvented.  On the mathematical side, the billiard regions and the reflection groups that emerge after compactification possess a rich structure which deserves investigation.

\subsection{Summary of Results}

The compactification analysis was recently carried out for eleven-dimensional supergravity and heterotic supergravity in \cite{Wesley:2006cd}. This showed that, in most cases, the BKL-limit does not produce a dominant wall set with the properties required of a valid Kac-Moody root system.
In this paper, we extend the investigation initiated in \cite{Wesley:2006cd} by studying compactifications of all the maximally oxidised theories originally classified by Breitenlohner, Maison and Gibbons \cite{Breitenlohner2} and further analysed in \cite{PopeJulia,Arjan}. These theories associate an Einstein-$p$-form system with any algebra $\mf{u}$ in the $A-G$ Cartan classification.  The Einstein-$p$-form system, when compactified on a torus, yields a scalar coset model in $D=3$ spacetime dimensions with U-duality symmetry algebra given by a split real form of $\mf{u}$, and whose BKL billiard is associated with the overextension $\mf{u}^{++}$ of this split real form. We perform an exhaustive analysis of all compactifications (defined by their vanishing Betti numbers) of all of the $A-G$ Einstein-$p$-form systems.  We present the Dynkin diagrams or Coxeter graphs describing the resulting billiards, and determine whether chaos is present in all possible cases.

We also elucidate the structure of the group of reflections in the dominant billiard walls that survive compactification, and its relation with the billiard region. The group is always a Lorentzian Coxeter group but the billiard region is not necessarily a fundamental domain of this Coxeter group. Rather it may correspond to a union of images of the fundamental Weyl chamber of the hyperbolic algebra $\mf{u}^{++}$ associated with the uncompactified theory. Using techniques from the theory of buildings we show that this region has a mathematical description as a sequence of pairwise adjacent chambers, known as a \emph{gallery}, inside the Cartan subalgebra $\mf{h}\subset\mf{u}^{++}$. Moreover, the resulting Coxeter group is then described by a non-standard presentation, with additional relations between the fundamental reflections, apart from the standard Coxeter relations. The Coxeter group itself might not be a simplex Coxeter group (i.e., a Coxeter group with fundamental region that is a simplex).  It is only when the Coxeter group is a simplex Coxeter group, and the billiard region is a fundamental domain, that there is a correspondence between billiard walls and simple roots of a Kac-Moody algebra.  This situation is generically absent in the compact setting (in the intermediate regime considered here).


\subsection{Outline of the Paper}

Our paper is organized as follows. Section \ref{section:BKL} gives a quick review of crucial elements of the billiard construction, with emphasis on the associated Coxeter group structure and its underlying geometric properties. In particular, we analyze some properties of convex polyhedra in hyperbolic space and explain how they can be used to understand certain features of Coxeter groups. In Section \ref{section:GeneralResults} we give some useful facts regarding the wall systems after compactification. We extend the set of theorems introduced and proven in \cite{Sophie} regarding the set of dominant walls for a given Einstein-dilaton-$p$-form system, to include the cases where individual $p$-form components are eliminated. We introduce the notions of \emph{formal Coxeter group} and \emph{billiard group} that will play important roles in subsequent developments. These concepts enable us to discuss general features of the Coxeter group structure of the reduced billiard, and we analyze in detail the properties of the new billiard domain. Most importantly, we show that this domain corresponds to a gallery, and we demonstrate how the structure of this gallery is related to the presentation of the associated Coxeter group.  In Section \ref{section:Examples} we describe in detail several especially illuminating cases in order to illustrate the techniques that we use, and to put the abstract results of Section \ref{section:GeneralResults} on a concrete footing.  In Section \ref{section:Conclusions} we end with concluding remarks and suggestions for future research.  Finally, a complete classification of all possible compactifications of maximally oxidised theories is given in the appendix.


\section{Coxeter Billiards and Compactification}
\label{section:BKL}
In this section we set the stage for the analysis carried out in Sections \ref{section:GeneralResults} and \ref{section:Examples}. We give a brief review of the billiard interpretation of the dynamics in the BKL-limit, without attempting completeness; more details can be found in a number of review articles \cite{Misner2,ArithmeticalChaos,DHNJ,DHNReview}. Following \cite{Wesley:2005bd,Wesley:2006cd} we also describe how compactification on manifolds of non-trivial topology modifies the billiard dynamics by projecting out walls associated with $p$-form fields which are rendered massive in the compactification. Furthermore, this section introduces some useful technology in the theory of Coxeter groups that will be crucial for understanding the main results of Section \ref{section:GeneralResults}. In particular, we analyze certain properties of convex polyhedra in hyperbolic space, and the associated Gram matrices.

\subsection{Cosmological Billiards and Coxeter Groups}
\subsubsection{The BKL-limit, Billiards and Kac-Moody Algebras}
Following Belinskii, Khalatnikov and Lifshitz we suppose the dynamics of gravity close to a spacelike singularity can be described, at each spatial point, as a piecewise linear motion of an auxiliary particle in a region of hyperbolic space \cite{BKL1,Chitre,Misner2}. The BKL analysis uses an asymptotic spacetime metric of the form
\beq
ds^{2}=-(g \, d\tau)^{2}+\sum_{i=1}^{d} e^{-2\be^{i}(x, \tau)} (\om^i)^2,
\eqnlab{BKLMetric}
\eeq
where $g$ is the determinant of the metric on spatial slices and $\om^{i}$ is the spatial ``Iwasawa frame''  (see \cite{DHNReview,LivingReview}). The singularity occurs at  $\tau\rightarrow \infty$ when the spatial volume density $g$ collapses locally at each spatial point. The equations of motion imply that the dynamics of the gravity-dilaton-$p$-form system at a fixed spatial point is that of a massless particle with coordinates $\be^{\mu}(\tau)$ moving in an auxiliary space, called \emph{$\beta$-space}, $\mf{M}_{\be}$, with metric
\beq
d\sigma^{2} =G_{\mu\nu}d\be^{\mu}d\be^{\nu}=\sum_i {d\be}^{i} d\be^{i}-\sum_{i,j}{d\be}^{i}{d\be}^{j} + \sum_{k=1}^{q}d\phi^{k} d\phi^{k}.
\eqnlab{MetricBetaSpace}
\eeq
This metric is flat and of Lorentzian signature, and so we have
\beq
\mf{M}_{\be}\simeq \mbb{R}^{1,M-1},
\eeq
where $M = d + q$ ($q$ being the number of dilatons $\phi$). The dynamics in the BKL-limit corresponds to free linear motion in $\mf{M}_{\be}$, interrupted by specular reflections against the walls defined by
$\om_{A}(\be)=0, $
where $\om_{A}\in\mf{M}_{\be}^{\star}$ are the wall forms. The wall forms include the symmetry walls
\beq
s_{ij}(\be)=\be^{i}-\be^{j},\quad (i, j=1, \dots, d \hs ;\hs i>j),
\eqnlab{SymmetryWalls}
\eeq
the gravitational walls\footnote{There are actually two types of gravity wall forms; the other set of wall forms being $G_i(\be)=\sum_{j\neq i}\be^{j}$. These are special because they are lightlike, $(G_i|G_i)=0$, in contrast with all the other wall forms which are spacelike. For the purposes of this paper the walls defined by $G_i(\be)=0$ are of no relevance since they are always subdominant with respect to the spacelike gravity walls defined by $G_{ijk}(\be)=0$.}
\beq
G_{ijk}(\be) = 2\beta^{i} + \sum_{m \ne i,j,k} \beta^m, \quad (i\ne j, \; i\ne k, \; j \ne k),
\eqnlab{GravityWalls}
\eeq
and, if $p$-form fields are present, they produce electric and magnetic walls
\beqa
{}e_{i_1 \cdots i_p}^{[p]}(\be) &=& \beta^{i_1} + \cdots + \beta^{i_p} +
\frac{1}{2} \lambda(p) \phi, \quad \hs \hs\hs\qquad
(i_1 < \cdots < i_p),
\nn \\
{}m_{i_1\cdots i_{d-p-1}}^{[p]}(\be) &=& \beta^{i_1} + \cdots + \beta^{i_{d-p-1}} -
\frac{1}{2} \lambda(p) \phi, \qquad
(i_1 < \cdots < i_{d-p-1}),
\eqnlab{ElectricMagneticWalls}
\eqa
where $\lambda(p)$ is the dilaton coupling for the $p$-form in question.  All of the wall forms presented above are spacelike, $(\om|\om)>0$, implying that the associated walls defined by $\om(\be)=0$ are all timelike.  Since the wall forms belong to the dual space $\mf{M}_{\be}^{\star}$, their scalar products are computed with the inverse metric
\beq
(\om|\om')=G^{\mu\nu}\om_{\mu}\om_{ \nu }'=\sum_i\om_{i}\om_{i}'-\f{1}{d-1}\big(\sum_{i} \om_{ i}\big)\big(\sum_j \om_{j}'\big)+\sum_{k=1}^{q}\om_{\phi^{k}}\om_{\phi^{k}}',
\eeq

The walls confine the billiard motion to the region $P_\beta$ where all the wall forms present in the case at hand obey $\om_A(\be)\geq 0$.  These wall forms can be expressed as linear combinations with nonnegative coefficients of a finite number $N$ of ``dominant" wall forms $\om_s$; hence the billiard region $P_{\be}$ is defined by:
\beq P_{\be}=  \big\{\be\in\mf{M}_{\be}\hs \big|\hs \om_{s}(\be)\geq 0, \hs s=1, \dots , N \big\}. \eeq The other (subdominant) walls are hidden behind the dominant ones, implying that they are irrelevant in the BKL-limit.

For a given theory, the precise structure of $P_\beta$ depends on the topology of spacetime because the relevant collection of $p$-form walls depends itself on the cohomology of the internal manifold in the pre-quantum gravity regime considered here \cite{Wesley:2005bd}. This will be discussed further in Section \ref{section:Compactification}.

Because the billiard walls are timelike hyperplanes in $\mbb{R}^{1, M-1}$ the reflections with respect to the walls preserve the future lightcone $\mc{L}^{+}\subset\mbb{R}^{1, M-1}$, and in particular they preserve the set of future-oriented, norm $-1$ vectors in $\mbb{R}^{1, M-1}$, corresponding to the hyperbolic space $\mbb{H}_m\subset\mc{L}^{+}$ ($M = m+1$).

Not all of the  $\be^{\mu}$ are physical because of the Hamiltonian constraint
\beq
\sum_i \pa_{\tau}\be^{i} \pa_{\tau}\be^{i}-\sum_{i,j}{\pa_{\tau}\be}^{i}{\pa_{\tau}\be}^{j}+\sum_{k=1}^{q}\pa_{\tau}\phi^{k}\pa_{\tau}\phi^{k}= 0.
\eeq
This constraint enables us to project onto a subspace of physical dynamical variables, which can be conveniently taken to be precisely the $m$-dimensional upper sheet of the unit hyperboloid just introduced
\beq
\mbb{H}_m=\{\be\in \mf{M}_{\be}\hs |\hs (\be |\be)=-1 \hbox{ and } \be \in \mc{L}^+\}.
\eeq
The lightlike linear motion in $\mf{M}_{\be}$ then corresponds to geodesics in hyperbolic space.  Similarly the walls, being timelike and going through the origin, project radially onto hyperplanes in $\mbb{H}_m$ (which are at the same time the intersections of the walls with $\mbb{H}_m$). The region in $\mbb{H}_m$ bounded by the (projected) hyperplanes will henceforth be referred to as the \emph{billiard table}.  It is customary to go back and forth between the projected description and the original description in Lorentzian space without always making the distinction explicit.

\subsubsection{The ``Uncompactified Billiard"}

Consider first the case when all the walls are switched on, as it occurs when no internal dimension is compact \cite{Wesley:2005bd}.  This is also the case relevant when the BKL analysis is pushed all the way to the singularity.  The billiard region is then the smallest possible one in the sense that the billiard region of all the other cases will contain the billiard region of the uncompactified theory. We shall denote by $\om_{A^{\p}}$ the dominant walls of the uncompactified theory. While the number of dominant walls relevant to the compact cases might not be equal to the dimension $M$ of $\mf{M}_{\be}$, it turns out that for all theories whose dimensional reduction to three dimensions is described by a symmetric space, the number of dominant walls is equal to $M$ \cite{ArithmeticalChaos,Sophie,DHNReview}.

In this case the dominant walls confine the billiard motion to the region $\mc{B}_{\be}\subset \mf{M}_{\be}$ defined by
\beq
\mc{B}_{\be}=\big\{\be\in\mf{M}_{\be}\hs \big|\hs \om_{A^{\p}}(\be)\geq 0, \hs A^{\p}=1, \cdots , M\big\}, \label{originalbilliard}
\eeq
The billiard table is a simplex in $\mbb{H}_m$. We shall call somewhat improperly the region (\ref{originalbilliard}) the ``uncompactified billiard region" and its projection on $\mbb{H}_m$ the ``uncompactified billiard table".

The scalar products $(\om_{A^{\prime}}|\om_{B^{\prime}})$ between the dominant walls can be organized into a matrix,
\beq
A_{A^{\prime}B^{\prime}}=\f{2(\om_{A^{\prime}}|\om_{B^{\prime}})}{(\om_{A^{\prime}}|\om_{A^{\prime}})}.
\eeq
In the noncompact case, the matrix $A$  turns out to possess the properties of a Lorentzian Cartan matrix \cite{ArithmeticalChaos,DHNReview}, thereby identifying the dominant wall forms $\om_{A^{\prime}}$ with the simple roots of the Kac-Moody algebra $\mf{g}(A)$ constructed from  $A$ \cite{Kac}. This Kac-Moody algebra is the ``overextension'' $\mf{u}^{++}$ of the U-duality algebra $\mf{u}$ \cite{Sophie}.  The group generated by the reflections in the billiard walls of the uncompactified theory is a Coxeter group, which is the Weyl group of the corresponding Kac-Moody algebra \cite{ArithmeticalChaos,DHNReview}.  We shall denote it by $\mf{W}$.

The action of $\mf{W}$ on $\mc{L}^{+}$ splits up into a disjoint union of chambers, called \emph{Weyl chambers}. One of these chambers is defined by the inequalities $\omega_{A^\prime} \geq 0$ and is called the \emph{fundamental chamber} $\mc{F}$. Then, all other chambers in $\mc{L}^{+}$ correspond to images of $\mc{F}$ under $\mf{W}$. The action of $\mf{W}$ on the Weyl chambers is simply transitive. When projected onto $\mbb{H}_m$, these chambers become $m$-simplices of finite volume. The fundamental chamber $\mc{F}$ is the uncompactified billiard region in which the chaotic dynamics takes place. The $m+1$ hyperplanes (or dominant walls) which bound the fundamental chamber, correspond to the codimension-one faces of $\mc{F}$ when projected onto $\mbb{H}_n$ (see also Section \ref{section:chambers}).

\subsection{Compactification and Cohomology}
\label{section:Compactification}
Compactification can modify the billiard, as was shown in \cite{Wesley:2005bd,Wesley:2006cd}. This occurs because the billiard dynamics in the intermediate regime considered here depends not only on the
$p$--form menu, but also on the topology of the space upon which the theory is
formulated, specifically on the de Rham cohomology $\mc{H}^{p}(\mc{M})$ of the compactification manifold $\mc{M}$. The rules for constructing the noncompact billiard system are given above, and here we focus on the ``selection rule" that describes how this billiard is modified after compactification.

\subsubsection{Selection Rule}

We study situations in which all spatial dimensions are compact, and thus spacetime $\Sigma$ has topology
\beq
\Sigma = \mbb{R}\times\mc{M},
\eeq
where $\mc{M}$ is closed, compact, and orientable. Electric and magnetic walls, $e(\beta)$ and $m(\be)$, arise from the electric and magnetic components, $F_E$ and $F_M$, of a given $p$-form $F$. On a compact manifold $\mc{M}$ the $p$-form fields which remain massless during the compactification correspond to solutions of the equations of motion of the form
\beq
F_E = f_E(t)\, \omega_p \wedge dt, \qquad
F_M = f_B(t)\, \omega_{p+1},
\eeq
where $\omega_p$ and $\omega_{p+1}$ are representatives of the de Rham cohomology classes $\mc{H}^p(\mc{M})$ and $\mc{H}^{p+1}(\mc{M})$, respectively. When $\mc{M}$ is compact, solutions that yield electric and magnetic billiard walls can therefore only be found when the de Rham cohomology classes are nontrivial.\footnote{For more general compactifications with fluxes turned on, this simple argument does no longer hold because then the massless spectrum is not determined only by the de Rham cohomology of $\mc{M}$. Instead one must consider a more complicated ``twisted'' cohomology (see, e.g., \cite{Gualtieri,Grana,Jarah}). However, in our analysis, these kinds of compactifications are actually trivial, since all $p$-forms are lifted and become massive. Hence, the billiard dynamics will be controlled by the purely gravitational sector, or the gravity-dilaton sector if dilatons are present. We thank Larus Thorlacius and Alexander Wijns for raising this issue, and for helpful discussions.}

We may now state the influence of the topology of $\mc{M}$ on the billiard structure simply in terms of a \emph{selection rule}. This rule makes use of the Betti numbers $b_j(\mc{M})$, which are the dimensions of the de Rham cohomology classes $\mc{H}^p(\mc{M})$. The selection rule reads \cite{Wesley:2005bd,Wesley:2006cd}:

\begin{itemize}
\item {\bf Selection Rule}: \emph{When $b_s (\mc{M})= 0$ for some $s$, we remove all billiard walls corresponding to electric $s$--forms, or  magnetic $(s-1)$--forms.}
\end{itemize}

\noindent  The selection rule is established using the same assumptions as the noncompact BKL analysis, namely that we are in a regime where classical gravity is valid, and studying a sufficiently ``generic" spacetime solution.

It has been known for some time that the algebraic structure of the billiard is invariant under Kaluza--Klein reduction on tori \cite{Sophie}. The selection rule is compatible with these results, since tori have no vanishing Betti numbers. Note also that none of the symmetry walls (or gravity walls) is eliminated by the compactification.  Hence, among the dominant walls of the compactified theory we always have the $(d-1)$ dominant symmetry walls $\beta^2 - \beta^1$, ..., $\beta^d - \beta^{d-1}$.

\subsection{Gram Matrices and Coxeter Groups}
In order to understand the structure of the reflection group that emerges when some $p$-form walls are switched off as well as the features of the corresponding billiard domain, it is useful to recall a few facts about polyhedra and reflection groups in hyperbolic space. The main reference for this section is \cite{Vinberg}.

\subsubsection{Convex Polyhedra}

We shall consider convex polyhedra $P$ of hyperbolic space, i.e., regions of the form  \beq P = \bigcap_{s = 1}^N H_s^+,\eeq where  $H_s^+$ is a half-space bounded by the hyperplane $H_s$, and $N$ is the number of such bounding hyperplanes.  In our case, $H_s$ is one of the walls of the relevant dominant wall system,
\beq
H_s=\{\be\in\mf{M}_{\be}\hs |\hs \om_s(\be)=0\},
\eeq
and $H_s^+$ is defined by
\beq
H_s^{+}=\{\be\in\mf{M}_{\be}\hs |\hs \om_s(\be)\geq 0\}.
\eeq
The polyhedra $P$ therefore contain the fundamental domain (\ref{originalbilliard}), defined by the dominant walls of the uncompactified theory. Hence it is clear that $P$ has non-vanishing volume.

\subsubsection{Relative Positions of Walls in Hyperbolic Space}
It is customary to associate with the convex polyhedron $P$ a matrix $G(P)$, the so-called \emph{Gram matrix}, which differs from the matrix $A$ by normalization. The construction proceeds as follows. To each hyperplane $H_s$ we associate a unit spacelike vector $e_s$ pointing inside $P$, i.e., pointing towards the billiard region (which is thus defined by $(\beta , e_s) \geq 0$).  We then construct the $N \times N$ matrix $G(P)$ of scalar products $(e_s | e_{s'})$. Four cases can occur for the scalar product $(e_s | e_{s'})$ between a given pair of distinct unit vectors $e_s$ and $e_{s'}$ \cite{Vinberg}\footnote{Note that one cannot have $e_s = - e_{s'}$ since then the region $H_s^+ \cap H_{s'}^+ = H_s$ has vanishing volume, which is excluded as we observed above.}: \begin{enumerate} \item $-1  \leq (e_s| e_{s'}) \leq 0. $ In this case, the hyperplanes $H_s$ and $H_{s'}$ intersect and form an acute angle. The limiting case $(e_s| e_{s'}) = -1$ means that the hyperplanes intersect at infinity, i.e., are parallel.  The other limiting case $(e_s| e_{s'}) = 0$ means that the hyperplanes form a right angle, which is both acute and obtuse.  \item $0  \leq (e_s| e_{s'}) \leq 1. $ In this case, the hyperplanes also intersect, but form an obtuse angle. The limiting case $(e_s| e_{s'}) = 1$ corresponds again to parallel hyperplanes meeting at infinity. \item $(e_s| e_{s'}) < -1$.  \item $(e_s| e_{s'}) >\phantom{-} 1$.   \end{enumerate} In the latter two cases, the hyperplanes diverge, i.e., do not meet even at infinity.  The fourth case and the limiting case $(e_s| e_{s'}) = 1$ ($s \neq s'$) are excluded from our analysis, because the geometry defined by the relevant electromagnetic walls has a metric which is positive semi-definite, so that one can have $(e_s| e_{s'}) \geq 1$ only if one of the walls is the (dominant) gravitational wall.  But this cannot arise, as shown by our analysis of scalar products involving the gravitational wall in Section \ref{section:GeneralResults} (taking into account the normalisation of the simple roots). 

We denote the dihedral angle between the hyperplanes $H_s$ and $H_{s^{\prime}}$ by $H_s^{+}\cap H_{s^{\prime}}^{+}$.  When the hyperplanes intersect, the value of the dihedral angle $H_s^+ \cap H_{s'}^+$ can be found from the relation \beq \cos (H_s^{+} \cap H_{s'}^{+}) = - (e_s| e_{s'}). \eeq

\subsubsection{Acute-Angled Polyhedra}
If the number of dominant walls is strictly smaller than the dimension $M$ of $\mf{M}_{\be}$, the billiard table in $\mbb{H}_m$ has infinite volume and the motion is non-chaotic.  After a finite number of collisions, the billiard ball escapes to infinity \cite{Demaret:1986ys,Damour:2000th}. We shall therefore assume that the number of dominant walls is greater than or equal to the dimension $M$ of $\mf{M}_{\be}$, a case that needs a more detailed analysis.  The Gram matrix is then of rank $M$ because among the $S$ dominant wall forms, one can find a subset of $M$ of them that defines a basis, namely the $(d-1)$ symmetry walls and one of the other dominant walls if there is no dilaton (or two linearly independent dominant non-symmetry walls if there is a dilaton etc.).  The convex polyhedron defined by the dominant walls is therefore non-degenerate (see \cite{Vinberg}). We shall also assume that the Gram matrix is indecomposable (cannot be written as a direct sum upon reordering of the $e_i$'s), as the decomposable case can be analysed in terms of the indecomposable one.

If the number of dominant walls is exactly equal to $M$, the billiard table is a simplex in hyperbolic space.  Otherwise, one has a non-simplex billiard table, with the number of faces exceeding dim $\mbb{H}_n+1=n+1$.

A crucial notion in the study of reflection groups is that of \emph{acute-angled polyhedra}.  A convex polyhedron is said to be {\em acute-angled} if for any pair of distinct hyperplanes defining it, either the hyperplanes do not intersect, or, if they do, the dihedral angle $H_s^+ \cap H_{s'}^+$ does not exceed $\frac{\pi}{2}$.  The Gram matrix, which has 1's on the diagonal, has then negative entries off the diagonal.  While non-degenerate, indecomposable acute-angled polyhedra on the sphere or on the plane are necessarily simplices, this is not the case on hyperbolic space.

\subsubsection{Coxeter Polyhedra and the Billiard Group}
\label{section:CoxeterPolyhedra}

We have seen that the dynamics of gravity is described in all cases (uncompactified or compactified) by the motion of a billiard ball in a region of hyperbolic space.  The reflections $s_s$ ($s=1, \dots, N$) with respect to the billiard walls generate a discrete reflection group which we want to characterize.  This group, which we shall call the \emph{billiard group} and denote by $\mf{B}$, is a subgroup of the Coxeter group relevant to the uncompactified case, where the total number of walls is maximum and the billiard region the smallest (and contained in all other billiard regions).  The billiard group is a crystallographic Coxeter group since it is generated by reflections \cite{Vinberg} and since it preserves the root lattice of the Kac-Moody algebra of the uncompactified case.  Its presentation in terms of the billiard walls might, however, be non-standard. The billiard group $\mf{B}$ will be examined more carefully in Sections \ref{section:BilliardGroup} and \ref{section:NonstandardPresentations}.

The billiard table has an important property which it inherits from the complete wall system of the theory. Consider the dihedral angle $H_i^+ \cap H_{j}^+$ between two different walls $H_i$ and $H_j$ that intersect.  If this angle is acute, then it is an integer submultiple of $\pi$, i.e., of the form $\pi/m_{ij}$ where $m_{ij}$ is an integer greater than or equal to $2$.  If this angle is obtuse, then $\pi - H_i^+ \cap H_{j}^+$ is an integer submultiple of $\pi$, i.e., $\pi - H_i^+ \cap H_{j}^+ = \pi/m_{ij}$, where $m_{ij}\in \mbb{Z}_{\geq 2}$. If the walls do not intersect, they are parallel and one has $m_{ij} = \infty$. Thus, given any pair of distinct walls, one can associate to it an integer $m_{ij} = m_{ji} \geq 2$.

In the case when all the angles are acute, and hence integer submultiples of $\pi$, the polyhedron is called a \emph{Coxeter polyhedron}.  Coxeter polyhedra are thus acute-angled.  In hyperbolic space, they may or may not be simplices.

\section{General Results}
\label{section:GeneralResults}

In this section we describe our new results concerning general features of the billiard structures after compactification.  In Section \ref{section:rules}, we use features of the wall system to explain why the billiard table need not be a Coxeter polyhedron after compactification. In Section \ref{section:FundamentalDomains} we describe the billiard region after compactification in terms of galleries, which we explain. Finally, we describe in Section \ref{section:ChaoticProperties} our methods for determining the chaotic properties for all possible compactifications.

\subsection{Rules of the Game}
\label{section:rules}

We show why the billiard region need not be a Coxeter polyhedron after compactification with the aid of two facts about billiard wall systems:

\begin{itemize}

\item \textbf{Fact 1}: \emph{If both an electric and a magnetic wall are present for
a given $p$--form, then the gravitational walls are subdominant.}  This was proven in \cite{Subcritical,Sophie}, by noticing that for any $p$ we have
\beq
G_{ijk} = e^{[p]}_{r_1 \cdots r_p} + m_{s_1\cdots s_{d-p-1}}^{[p]},
\eeq
where one of the $r_q$ and one of the $s_q$ are equal to $i$, and neither $j$ nor $k$ appears in either the $r_q$ or $s_q$.

\item \textbf{Fact 2}: \emph{The inner product between a gravitational wall and a $p$--form wall is unity for}
\begin{itemize}
\item \emph{electric walls with} $p \leq D-3$,
\item \emph{magnetic walls with} $p \geq 1$,
\end{itemize}
\emph{and the inner product vanishes for}
\begin{itemize}
\item \emph{electric walls with} $p=D-2$,
\item \emph{magnetic walls with} $p=0$.
\end{itemize}
Typically, $p$ forms with $p > \lfloor D/2 \rfloor - 1$ are dualised so that we may safely assume that $p \leq \lfloor D/2 \rfloor - 1$.  (This is a stronger condition than $p \leq D-3$ for $D \geq 4$).  Then, we have that the inner product between gravitational and $p$--form walls is always positive, except when the $p$--form is a scalar (axion) and the inner product vanishes.  This fact is proven by computing the relevant inner products (which are independent of the dilaton coupling(s) of the $p$--form) using the metric between wall forms.

This fact is significant in combination with the requirement that a system of dominant walls define an acute-angled polyhedron if the inner products between each pair of walls are either zero or negative. Therefore we can conclude:

\emph{If the dominant wall set includes a gravitational wall and any non-scalar $p$--form wall, then the dominant wall system does not define a Coxeter polyhedron.}

\end{itemize}
\noi For compactifications on tori, only {\bf Fact 1} is relevant since all components of the same $p$-forms are preserved in the compactification \cite{Sophie}. Hence, electric and magnetic walls are always present in pairs, ensuring that the gravity wall is always subdominant. More general compactifications can eliminate one of the electric or magnetic walls of a given $p$--form, while leaving the other intact.  Unlike the noncompact case, it is possible for both gravitational and $p$--form walls to appear simultaneously in the set of dominant walls. {\bf Fact 2} tells us that when this happens, we no longer have a Coxeter polyhedron.

In our analysis, we never eliminate gravitational walls (although there are some partial results regarding their selection rules that were given in \cite{Wesley:2005bd}).  This means that after compactification there is always a gravitational wall in the root system, though it is not necessarily dominant.  Therefore all compactifications we study fall into one of three classes:

\begin{itemize}

\item A pair of electric and magnetic walls from the noncompact theory has not been eliminated by compactification, so the gravitational wall is subdominant.  The resulting billiard table may be a Coxeter polyhedron, depending on the details of the $p$-form menu and couplings in the theory. (Example: the $b_1=0$ compactification of the $B_n$ sequence, see Section \ref{section:Examples}.)

\item One (or both) members of each pair of electric/magnetic walls are eliminated, so the gravitational wall is exposed.  If there are any other $p$-form walls left, because of {\bf Fact 2}, the
billiard table cannot be a Coxeter polyhedron.  (Example: the $b_3=0$ compactification in the $E_8$ sequence.)  However, it may occur that ``coincidentally" two walls from different $p$-forms can combine and make the gravitational wall subdominant.

\item All of the $p$-form walls are eliminated by compactification, and so only the gravitational wall is left.  In this case one always obtains $A_n^{++}$, the billiard corresponding to pure gravity.  This billiard sits at the bottom of every list when all possible Betti numbers are set to zero.  Occasionally there are also direct summands corresponding to scalar fields that are never eliminated by compactification. (Example: $b_1=b_2=0$ for $C_n$, with $n>4$, as a billiard described by $A_1^{++}\times C_{n-1}$.)

\end{itemize}

Only the first of the cases described above arises in the noncompact theory, where gravitational walls are never dominant, except when they are the only non-symmetry walls.

\subsection{Describing the Billiard Group After Compactification}
\label{section:BilliardGroupAfterCompactification}

We shall now begin to assemble the various structures described so far in order to analyze the group-theoretical properties of the billiard dynamics after compactification. This involves understanding the relation between the billiard table and the fundamental domain of the associated reflection group. In this context it is important to distinguish between the formal Coxeter group and the billiard group. We consider these concepts in turn in the following two sections, and explain how they are related in Section \ref{section:NonstandardPresentations}.

\subsubsection{The Formal Coxeter Group}
Recall from Section \ref{section:CoxeterPolyhedra} that the reflections $s_i$ with respect to the dominant walls $H_i$ generate a Coxeter group. To describe this group we must characterize the relations among the reflections $s_i$. Being reflections, they clearly satisfy \beq s_i^2 = 1 \label{Cox1}. \eeq  Consider next the reflections $s_i$ and $s_j$ with respect to two different hyperplanes $H_i$ and $H_j$. Then, the product $s_i s_j$ is a rotation by the angle $\frac{2 \pi}{m_{ij}}$ (where the integers $m_{ij}$ were introduced in Section \ref{section:CoxeterPolyhedra}) and so
\beq (s_i s_j)^{m_{ij}} = 1. \label{Cox2} \eeq

These relations alone define a Coxeter group, which we shall call the \emph{formal Coxeter group} $\mf{C}$ associated with the billiard. One can describe $\mf{C}$ more precisely as follows. Let $\tilde{\mf{C}}$ be the group freely generated by the elements of the set $\mathcal{S}=\{r_1, \dots, r_N\}$, and let $\mf{N}$ be the normal subgroup generated by $(r_i r_j)^{m_{ij}}$, where the \emph{Coxeter exponents} $m_{ij}$ satisfy \cite{Kac,Humphreys} (see also \cite{LivingReview})
\beqa
{}Êm_{ii}&=&1,
\nn \\
 {}Êm_{ij}&\in& \mathbb{Z}_{\geq 2}, \hs  i\neq j,
 \nn \\
{}Êm_{ij}&=& m_{ji}.
\eqa
Then $\mf{C}$ is defined as the quotient group $\tilde{\mf{C}}/\mf{N}$ and has the following standard presentation:
\beq
\mf{C} =  \left< r_1,\dots, r_n\hs |\hs (r_ir_j)^{m_{ij}}=1, \hs i,j=1,\dots, N \right>.
\eeq
One can associate a Coxeter graph with the formal Coxeter group, i.e., with the set of $m_{ij}$'s.  Each $r_i$ defines a node of the Coxeter graph, and two different nodes $i$ and $j$ are connected by a line whenever $m_{ij} >2$, with $m_{ij}$ explicitly written over the line whenever $m_{ij} >3$ \cite{Humphreys}. To a Coxeter graph, one can further associate a symmetric matrix defined by \beq B_{ij} = -\cos \Big(\f{\pi}{m_{ij}}\Big) \eeq with 1's on the diagonal and non-positive elements otherwise.

\subsubsection{The Billiard Group and its Fundamental Domain}
\label{section:BilliardGroup}
In Section \ref{section:CoxeterPolyhedra} we introduced the concept of the \emph{billiard group}Ê $\mf{B}$. Here we shall elaborate on this object and elucidate the structure of its fundamental domain.

The billiard group $\mf{B}$ is defined as the group generated by the reflections $s_i$ with respect to the dominant walls of the billiard table. This group coincides with the formal Coxeter group $\mf{C}$ (with $r_i \equiv s_i$) if and only if there are no additional relations among the $s_i$'s. Two cases must be considered.

\begin{enumerate}
\item {\bf Acute-angled billiard tables.}  There is no additional relations among the $s_i$'s if and only if the billiard table is acute-angled, i.e., is a Coxeter polyhedron \cite{Vinberg}.  In that case, the matrix $B=(B_{ij})$ coincides with the Gram matrix $G(P)$.  Furthermore, the billiard table is a fundamental domain \cite{Vinberg}.
    
    In the hyperbolic case relevant here, the billiard table need not be a simplex.  When it is a simplex, however, there is further structure.  The Gram matrix is non-degenerate and the action of the Coxeter group on the $\beta$-space $\mf{M}_{\be}$ coincides with the standard geometric realization considered in \cite{Humphreys}.  In addition, the matrix \beq
A_{ss^{\prime}}=\f{2(\om_{s}|\om_{s^{\prime}})}{(\om_{s}|\om_{s})},
\eeq has non-positive integers off the diagonal and hence is a non-degenerate Cartan matrix.  It is obvious that the off-diagonal entries are integers since the walls correspond to some roots of the Kac-Moody algebra of the uncompactified case. Moreover, it follows from the fact that the billiard table is acute-angled that they are negative. The billiard group is then the Weyl group of a simple Kac-Moody algebra.\footnote{In the case of a non-simplex, acute-angled billiard table, the matrix $A$ is also a valid Cartan matrix, but it is degenerate and the corresponding Kac-Moody algebra is not simple. Furthermore, the standard geometric realization is defined in a space of dimension larger than $M$, while the billiard realization is defined in the $M$-dimensional $\beta$-space.}  When the matrix $A$ is a Cartan matrix, one can associate to it a Dynkin diagram.

\item {\bf Non acute-angled billiard tables.} If the billiard table is not acute-angled (and hence not a Coxeter polyhedron), there are further relations among the $s_i$'s and the billiard group is therefore the quotient of the formal Coxeter group $\mf{C}$ by these additional relations.  Moreover, the billiard table is not a fundamental domain of the billiard group. One may understand this feature as follows.  Consider a dihedral angle $H_i^+ \cap H_{j}^+$ of the polyhedron that is obtuse.  The rotation $s_i s_j$ by the angle $2 \pi / m_{ij}$ is an element of the group and hence maps reflection hyperplanes to reflection hyperplanes. The image by this rotation of $H_j$ is the hyperplane $s_i H_j$ that intersects the interior of the billiard table.\footnote{When the Coxeter group is crystallographic ($m_{ij} \in \{2,3,4,6\}$), the converse is also true: if the angle between $H_i$ and $H_j$ is acute, then the image $s_iH_j$ does \emph{not} intersect the interior of the billiard table.} The reflection $s_i s_j s_i$ with respect to this hyperplane belongs to the group, and hence the billiard table cannot be a fundamental region of the billiard group since the orbit of a point sufficiently close to $s_i H_j$ intersects the billiard table at least twice.  Fundamental regions are obtained by considering all the mirrors (reflection hyperplanes) associated with the group (most of which are not billiard walls), which decompose the space into equivalent chambers that are permuted by the group (homogeneous decomposition). Each of these chambers is a fundamental domain. The billiard group $\mf{B}$ is generated by the reflections in the mirrors of the fundamental domain, which provide a standard presentation of the group, and the billiard table is a union of chambers \cite{Vinberg}. Examples of the occurrence of this phenomenon will be discussed below.

Although the billiard table is not a fundamental domain, it can be naturally described as a gallery defined by the Coxeter group of the uncompactified theory. This is described in Section \ref{section:BilliardGallery}.

\end{enumerate}

\subsubsection{Non-Standard Presentations of Coxeter Groups}
\label{section:NonstandardPresentations}

We have seen how one can associate a formal Coxeter group $\mf{C}$ to the billiard region using the Coxeter presentation $\mf{C}=\tilde{\mf{C}}/\mf{N}$. The billiard group $\mf{B}$ -- which is also a Coxeter group -- differs from $\mf{C}$ when the billiard table possesses obtuse angles because the reflections in the walls of the billiard then fulfill additional relations. This yields a non-standard presentation of the billiard group, which can be formally described as follows.

Let $\mc{B}_{\be}$ be the billiard region after compactification, let the elements of the set $\mc{S}=\{s_i\hs |\hs i=1, \dots, N\}$ be the reflections in the walls $W_i$ bounding $\mc{B}_{\be}$, and let $\tilde{\mf{C}}$ be the formal group freely generated by $\mc{S}$. The dihedral angles between the $W_i$ give rise to a set of Coxeter exponents $m_{ij}$, with associated normal subgroup $\mf{N}\subset \tilde{\mf{C}}$. Suppose now that the region $\mc{B}_{\be}$ is not a fundamental domain of $\mf{B}$, and denote by $\mf{J}$ the normal subgroup of $\tilde{\mf{C}}$ generated by $(s_is_j)^{m_{ij}}$ and any other non-standard relations between the elements of $\mc{S}$. Note that we have $\mf{N}\subset \mf{J}$. The billiard group $\mf{B}$ is then the quotient
\beq
\mf{B}=\tilde{\mf{C}} / \mf{J}.
\eeq
Equivalently, if we denote by $\mf{F}$ the normal subgroup of $\tilde{\mf{C}}$ generated only by the non-standard relations between the elements of $\mc{S}$, we may describe the billiard group as
\beq
\mf{B}=\mf{C}/\mf{F}.
\eeq
Neither of these presentations is a Coxeter presentation.

 In all cases we consider in this paper, the uncompactified billiard is described by the Weyl group $\mf{W}[\mf{u}^{++}]$ of a Lorentzian Kac-Moody algebra $\mf{u}^{++}$. For general compactifications, the billiard group $\mf{B}$ is therefore a Coxeter subgroup of $\mf{W}[\mf{u}^{++}]$. However, we shall also see examples of cases when the formal Coxeter group $\mf{C}$ after compactification differs from $\mf{u}^{++}$, while the billiard group $\mf{B}$ is actually isomorphic $\mf{u}^{++}$, with a non-standard presentation. This is described in detail in Section \ref{section:E7}.

\subsection{Fundamental Domains, Chamber Complexes and Galleries}
\label{section:FundamentalDomains}

We have seen that the billiard table need not be a Coxeter polyhedron upon compactification.  When it is not a Coxeter polyhedron, it no longer corresponds a fundamental domain of the billiard group $\mf{B}$. Moreover, $\mf{B}$ is the quotient by nontrivial extra relations of the formal Coxeter group associated with the billiard table. In this section, we describe how the billiard region relevant to the compactified case can then be built as a union of images of the uncompactified billiard region.  This is achieved using the theory of buildings, in terms of chambers and galleries.

\subsubsection{Chambers and Galleries}
\label{section:chambers}
The analysis in this section makes use of the treatment of Coxeter groups as a theory of \emph{buildings}, a formalism mainly developed by J. Tits. Introductions and references may be found in \cite{Garrett,Brown}.

The basic idea is to study Coxeter groups in geometric language by defining them in terms of the objects on which they act nicely. The \emph{buildings} are the fundamental objects which then defines the associated Coxeter group. For example, finite Coxeter groups act on so-called \emph{spherical buildings}, because these groups preserve the unit sphere. We are interested in Coxeter groups which act on \emph{hyperbolic buildings}, i.e., hyperbolic Coxeter groups, which preserve the hyperbolic space.

An $n$-\emph{simplex} $\mc{X}$ in hyperbolic space is determined by its $n+1$ vertices.  A $1$-simplex is determined by its two endpoints, a $2$-simplex (a triangle) is determined by its three vertices etc.  For this reason, it is convenient to identify an $n$-simplex $\mc{X}$ with the set of its $n+1$ vertices, $\mc{X}=\{\mbox{set of}\ n+1\ \mbox{vertices}\}$.  A \emph{face} $f$ of $\mc{X}$ is a simplex corresponding to a non-empty subset $f\subset \mc{X}$. The \emph{codimension} of $f$ with respect to $\mc{X}$ is given by $\text{dim}\ \mc{X}-\text{dim}\ f$. For example, there are three codimension-one faces in a 2-simplex, which are the three edges of the triangle.

\begin{figure}[ht]
\begin{center}
\includegraphics[width=100mm]{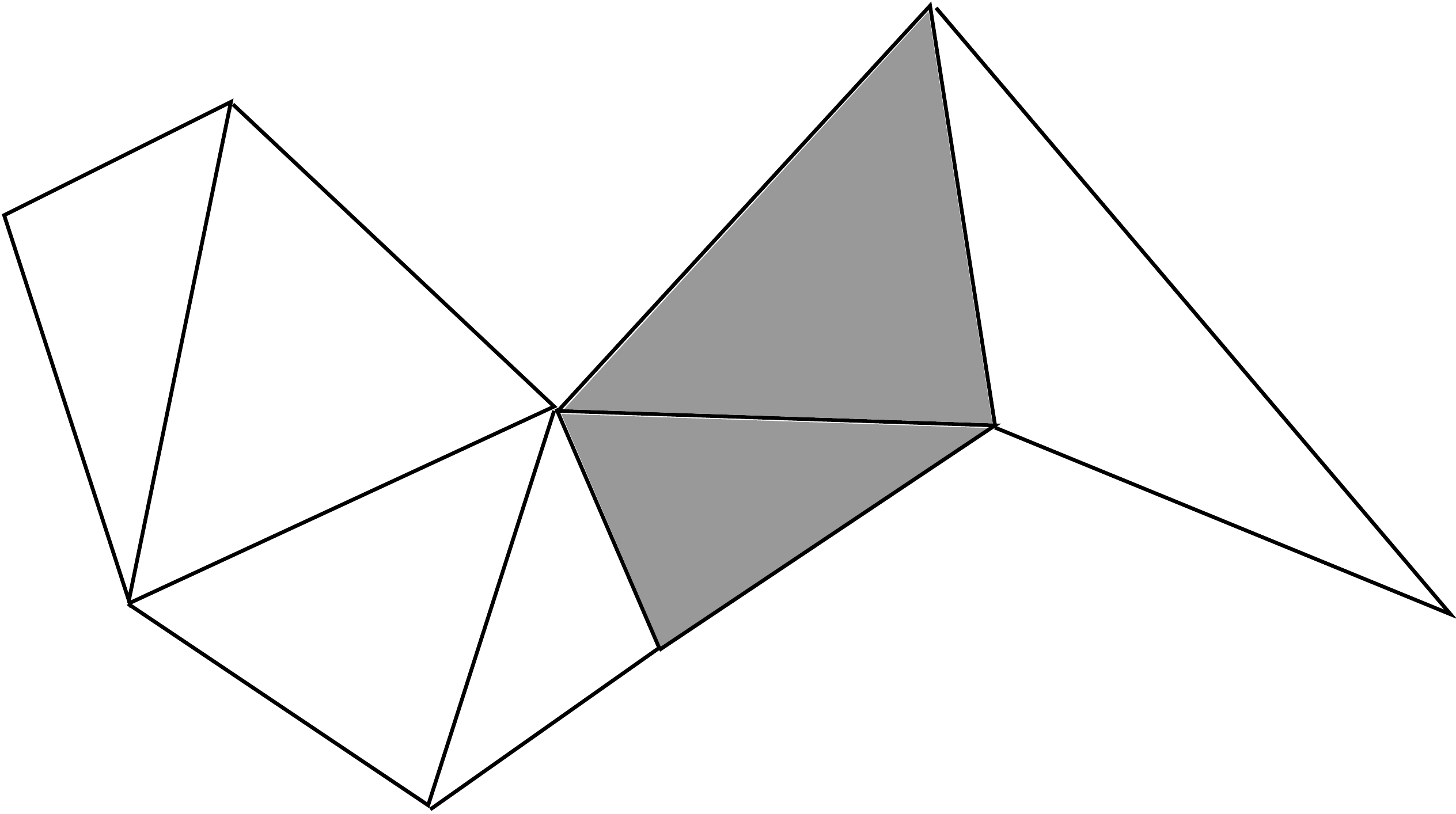}
\caption{A simplicial complex of $2$-simplices (triangles). The two shaded regions represent adjacent (maximal) simplices.}
\label{figure:Simplices}
\end{center}
\end{figure}

Next we define the notion of a simplicial complex. Let $V$ be a set of vertices, and $\mc{K}$ a set of finite subsets $\mc{X}_k\subset V$. We assume that the subsets containing a single vertex of $V$ are all elements of $\mc{K}$. Then $\mc{K}$ is called a \emph{simplicial complex} if it is such that given $\mc{X}\in \mc{K}$ and a face $f$ of $\mc{X}$, then $f\in \mc{K}$. The elements $\mc{X}_k$ of $\mc{K}$ are the simplices in the simplicial complex. Two simplices $\mc{X}_1$ and $\mc{X}_2$ of the same dimension in $\mc{K}$ are called \emph{adjacent} if they share a codimension-one face, i.e., if they are separated by a common wall. Figure \ref{figure:Simplices} illustrates a simplicial complex of $2$-simplices.

\begin{figure}[t]
\begin{center}
\includegraphics[width=140mm]{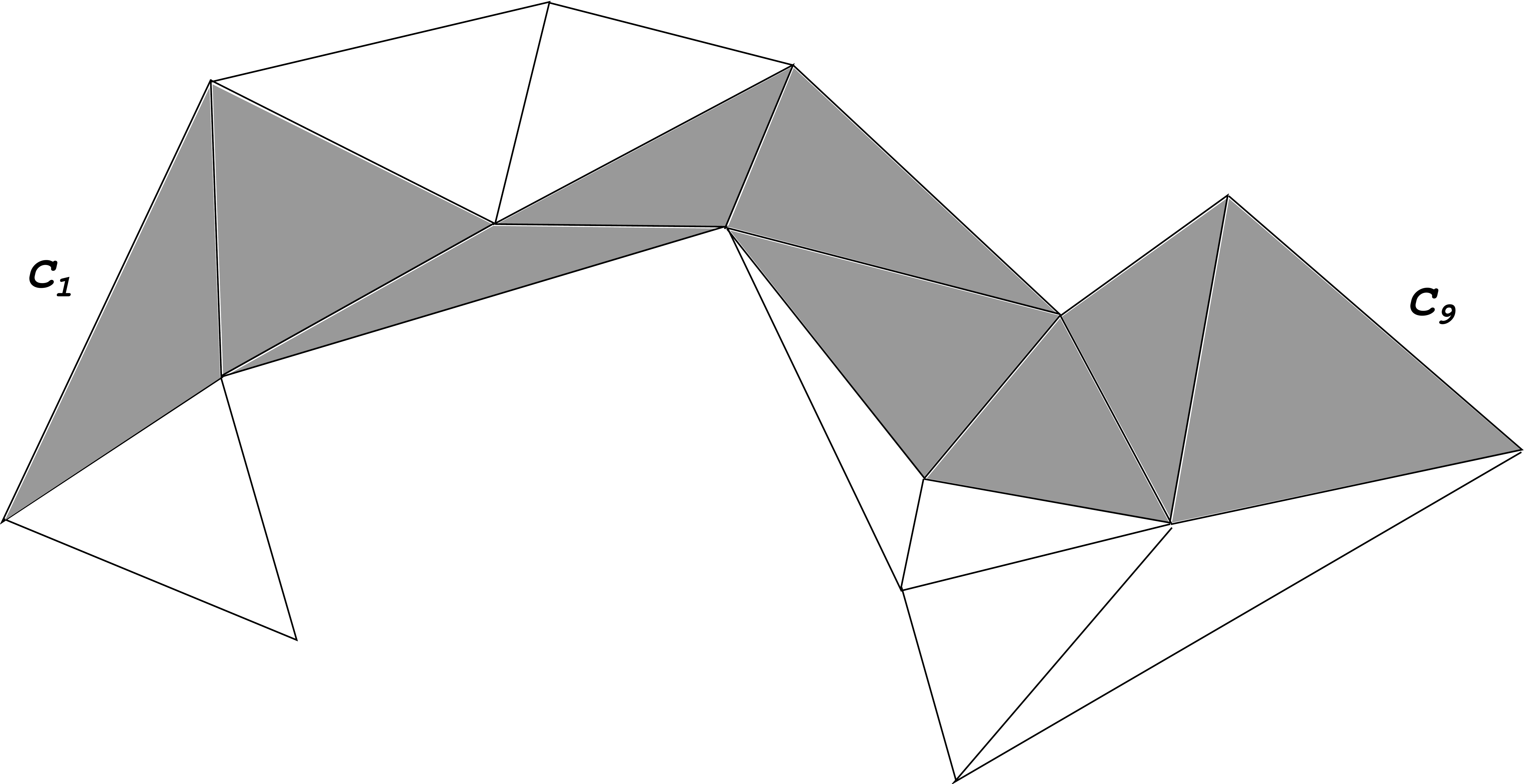}
\caption{A chamber complex with a gallery $\Gamma : \mc{C}_1,\dots, \mc{C}_9$, represented by the shaded region. The length of the gallery is $k=9$, which is also the distance between $\mc{C}_1$ and $\mc{C}_9$ since $\Gamma$ is the shortest gallery connecting $\mc{C}_1$ and $\mc{C}_9$.}
\label{figure:GalleryExample}
\end{center}
\end{figure}

A \emph{maximal simplex} $\mc{C}$ in $\mc{K}$ is such that $\mc{C}$ does not correspond to the face of another simplex in $\mc{K}$. Maximal simplices in a simplicial complex are called \emph{chambers} and they shall be our main objects of study. A sequence of chambers $ \mc{C}_1, \dots, \mc{C}_k$, such that any two consecutive chambers $\mc{C}_i$ and $\mc{C}_{i+1}$ are adjacent is called a \emph{gallery} $\Gamma$. Thus, a gallery corresponds to a connected path between two chambers $\mc{C}_1$ and $\mc{C}_k$ in $\mc{K}$, and we write
\beq
\Gamma \hs :\hs  \mc{C}_1, \mc{C}_2, \dots, \mc{C}_{k-1}, \mc{C}_k.
\eeq
The \emph{length} of $\Gamma$ is $k$, and the \emph{distance} between $\mc{C}_1$ and $\mc{C}_k$ is the length of the shortest gallery connecting them. If any two chambers in $\mc{K}$ are connected by a gallery, then the simplicial complex is called a \emph{chamber complex}. A simple example of a gallery inside a chamber complex is displayed in Figure \ref{figure:GalleryExample}.

\subsubsection{The Billiard Region as Gallery}
\label{section:BilliardGallery}

We describe the billiard regions of compactified gravity-dilaton-$p$-form theories in the language presented above. This is achieved by expressing them in terms of the billiard region of the uncompactified theory (or compactified on a torus) which we recall is the fundamental Weyl chamber $\mc{F}$ of the Weyl group $\mf{W}[\mf{u}^{++}]$ of the Kac-Moody algebra $\mf{u}^{++}$, whose Cartan matrix is defined through the scalar products between the dominant wall forms.

Compactification amounts to a process of removing dominant walls, and so the billiard table is enlarged.  The inequalities $\omega_{A^\prime} \geq 0$ associated with the simple roots of the underlying Kac-Moody algebra are indeed replaced by weaker inequalities.  The bigger region so defined can be written as a union of Weyl chambers of $\mf{W}[\mf{u}^{++}]$.  We shall illustrate this phenomenon on the example of the familiar Lie algebra $A_3$, whose Weyl group is a spherical Coxeter group. The fundamental Weyl chamber $\mc{F}$ is defined by \beq \om_1 (\be) \geq 0,\; \; \om_2 (\be) \geq 0 , \; \; \om_3 (\be) \geq 0, \eeq corresponding to the three simple roots.  The non simple roots are $\om_1 + \om_2$, $\om_2 + \om_3$ and $\om_1 + \om_2 + \om_3$.
Suppose that the single dominant wall $W_{1}$ defined by $\om_1(\be) = 0$ is suppressed. Effectively, this implies that the particle geodesic may cross the wall $W_{1}$. Thus it moves from the region where $\om_1(\be) \geq 0$ into the region where
\beq
\om_{1}(\be)\leq 0.
\eeq
We shall consider two cases. (i) The new billiard region is defined by \beq  \om_1 (\be) + \om_2 (\be) \geq 0,\; \; \om_2 (\be) \geq 0 , \; \; \om_3 (\be) \geq 0, \eeq i.e., the wall $W_1$ is replaced by the wall $\om_1 (\be) + \om_2 (\be) = 0$;  (ii) In the second case, we suppose that also the wall defined by $\om_1(\be)+\om_2(\be)=0$ is suppressed. Then the new billiard region is defined by the inequalities \beq \om_1 (\be) + \om_2 (\be) + \om_3(\be) \geq 0,\; \; \om_2 (\be) \geq 0 , \; \; \om_3 (\be) \geq 0 \eeq i.e., the wall $W_1$ is replaced by the wall $\om_1 (\be) + \om_2 (\be) + \om_3(\be) = 0$.

In the first case, one can write the new billiard region as the union $\mc{F} \cup \mc{A}_2$ where $\mc{A}_2$ is defined by  the inequalities \beq  \om_1 (\be) \leq 0,\; \; \om_1(\be) + \om_2 (\be) \geq 0 , \; \;  \om_3 (\be) \geq 0 \eeq (which imply $\om_2 (\be) \geq 0$). The region $\mc{A}_2$ is the Weyl chamber obtained by reflecting the fundamental Weyl chamber across the wall $W_1$ since
\beqa
{}s_1(\om_1)&=&-\om_1,
\nn \\
{}s_1(\om_2)&=&\om_1+\om_2,
\nn \\
{}s_1(\om_3)&=&\om_3.
\eqa
Hence,
\beq
s_1\cdot \mc{F}=\{\be \in\mf{h}\hs|\hs \om_1(\be)\leq 0, (\om_1+\om_2)(\be)\geq 0, \om_3(\be)\geq 0\} = \mc{A}_2.
\eeq
A reflection of this type which maps a chamber $\mc{C}$ to an adjacent chamber $\mc{C}^{\prime}$ is known as an \emph{adjacency transformation} \cite{Vinberg}.
By removing the single dominant $W_1$, we therefore get in the first case a new region which is precisely twice as large as the original fundamental region.

\begin{figure}[h]
\begin{center}
\includegraphics[width=70mm]{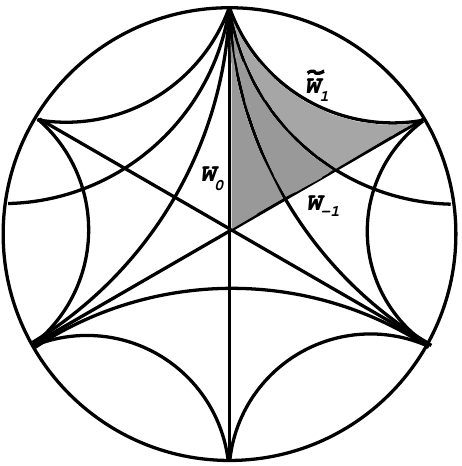}
\caption{Here we illustrate a gallery $\Gamma\hs :\hs \mc{C}_1, \mc{C}_2, \mc{C}_3$ of length 3 for the case of the hyperbolic Kac-Moody algebra $A_1^{++}$. The two walls $W_{0}$ and $W_{-1}$ are associated with the affine and overextended simple roots $\al_0$ and $\al_{-1}$, respectively. The original fundamental Weyl chamber $\mc{C}_1=\{h\in\mf{h}\ |\ \al_1(h)\geq 0, \hs \al_0(h)\geq 0, \hs \al_{-1}(h)\geq 0 \}$ corresponds to the leftmost shaded region. We have removed the wall $W_1=\{h\in\mf{h}\ |\ \al_1(h)=0\}$ as well as the wall $(\al_0+\al_1)(h)=0$. The far end of the billiard region is now bounded by the new wall $\tilde{W}_1=\{h\in\mf{h}\ |\ (2\al_0+3\al_1)(h)=0\}$. Each of the three chambers is clearly a copy of the fundamental chamber, and the total region is of finite volume. See, e.g., \cite{Feingold:2003es,LivingReview} for more detailed discussions of the Weyl group of $A_1^{++}$. }
\label{figure:A1++TilingGallery}
\end{center}
\end{figure}

In the second case, although we also remove a single dominant wall of the original billiard, we get a larger region.  This is because we also remove the subdominant wall $\om_1 (\be) + \om_2(\be) = 0$ which is exposed once $W_1$ is removed.  Indeed, the new billiard region can now be written as the union $\mc{F} \cup \mc{A}_2 \cup \mc{A}_3$ where $\mc{A}_3$ is defined by  the inequalities \beq \om_2(\be) \geq 0, \; \; \om_1(\be) + \om_2 (\be) \leq 0 , \; \;   \om_1(\be) + \om_2 (\be)+ \om_3 (\be) \geq 0 \eeq (which imply $\om_1 (\be) \leq 0$ and $\om_3 (\be) \geq 0$).  The region $\mc{A}_3$ is again a Weyl chamber, obtained from $\mc{A}_2$ by acting with the reflection $s$ with respect to $\om_1+\om_2$ since
\beqa
{}s(-\om_1)&=&\om_2
\nn \\
{}s(\om_1 + \om_2)&=&- (\om_1+\om_2),
\nn \\
{}s(\om_3)&=&\om_1 + \om_2 + \om_3.
\eqa
Hence,
\beq
s\cdot \mc{A}_2=\{\be \in\mf{h}\hs|\hs \om_2(\be)\geq 0, (\om_1+\om_2)(\be)\leq 0, \om_1(\be) + \om_2 (\be) + \om_3(\be)\geq 0\} = \mc{A}_3.
\eeq
Note that $\mc{A}_2$ and $\mc{A}_3$ are adjacent.  Thus, while in the second case we also remove a single dominant wall, we now obtain a region three times larger than the fundamental Weyl chamber. This new region can be described as a union of three pairwise adjacent Weyl chambers. In Figure \ref{figure:A1++TilingGallery} we describe pictorially a similar example for the case of the hyperbolic Coxeter group $A_1^{++}$.

By extrapolating this analysis to the general case where we remove an arbitrary number $r\leq n+1$ of dominant walls, we may conclude that the new billiard region corresponds to a union of images of the fundamental Weyl chamber. This naturally has the structure of a simplicial complex, and moreover, by a suitable ordering of the chambers, one sees that it corresponds to a gallery covering the whole complex.\footnote{A similar gallery-type structure was recently uncovered in a very different physical context in \cite{Miranda}.}  In conclusion, we have found the following: \emph{the billiard region $\mc{B}$ obtained by compactification on a manifold of non-trivial topology is described by a gallery $\Gamma$ inside the Cartan subalgebra $\mf{h}$ of the original hyperbolic Kac-Moody algebra $\mf{g}$.}

The dynamics after compactification is chaotic if the new billiard region is a finite union of images of the fundamental chamber, i.e., if the gallery $\Gamma$ has finite length, while if this union is infinite the particle motion will eventually settle down in a single asymptotic Kasner solution, and chaos is removed.  Since the Coxeter reflections preserve the volume, the volume of $\mc{B}$ is
\beq
\text{vol}\ \mc{B}=k\cdot \text{vol}\ \mc{F},
\eeq
where $k$ is the length of the gallery $\Gamma$ associated with $\mc{B}$.

\subsection{Determining the Chaotic Properties After Compactification}
\label{section:ChaoticProperties}

The selection rules described in Section \ref{section:BKL} provide a straightforward means to determine the billiard system after compactification.   Determining whether or not this billiard system is chaotic, i.e., computing the biliard table volume, is somewhat more involved because finding explicitly the corresponding galleries might be intricate.   In most cases it is possible to answer this question purely analytically without working out the gallery, although there are several different techniques that work for different types of billiard system. In this section we describe the various methods we employ.

\begin{figure}[ht]
\begin{center}
\includegraphics[width=150mm]{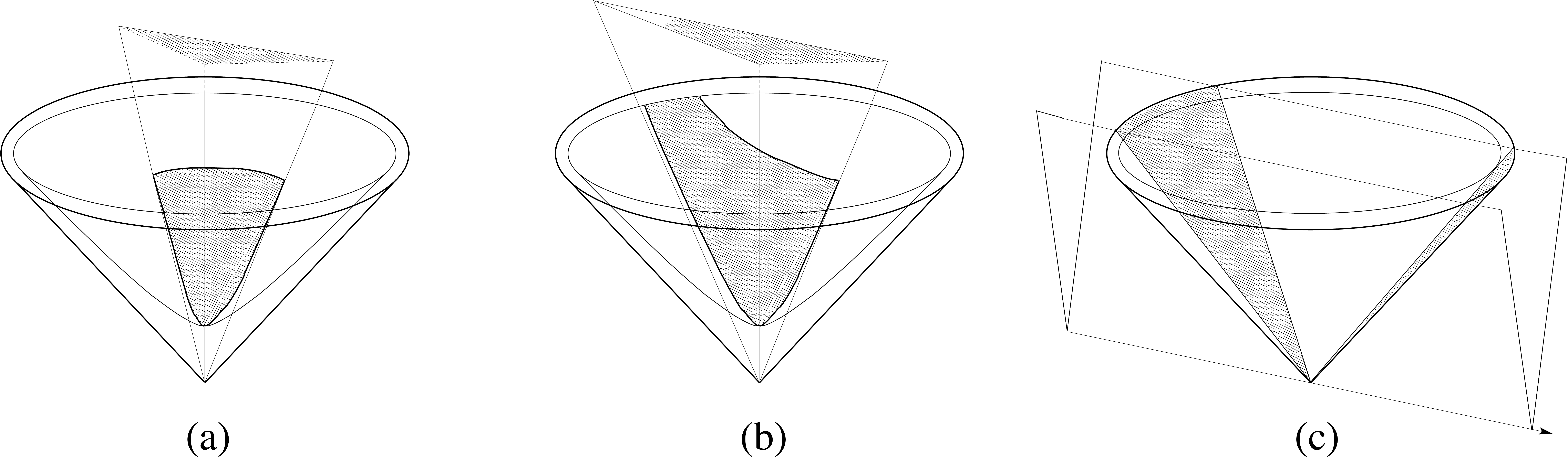}
\caption{Some examples of the wall systems and their chaotic properties: ({\bf a}) the wall system corresponding to a hyperbolic Kac-Moody algebra, ({\bf b}) the wall system of a non-hyperbolic Kac-Moody algebra, or one for which the coweight construction is possible, ({\bf c}) a wall system with fewer walls than the dimension of the $\beta$-space. }
\label{figure:WallCones}
\end{center}
\end{figure}

The simplest case is when the billiard table is a Coxeter simplex.  The matrix $\bar{A}_{ab}$ is then a Cartan matrix. The associated Kac-Moody algebra $\mf{g}(\bar{A})$ is by construction a regular subalgebra\footnote{A subalgebra $\bar{\mf{g}}\subset\mf{g}$ is \emph{regularly embedded} in $\mf{g}$ if and only if two conditions are
fulfilled: (i) the root vectors of $\bar{\mf{g}}$ are root vectors of
$\mf{g}$; and (ii) the simple roots of $\bar{\mf{g}}$ are real roots of
$\mf{g}$. Moreover, the embedding is \emph{positive regular} if the positive root vectors of $\bar{\mf{g}}$ are positive root vectors of $\mf{g}$. See, e.g., \cite{Feingold:2003es,GeometricConf} for more detailed discussions on regular subalgebras of Kac-Moody algebras.} of the Kac-Moody algebra $\mf{g}(A)$, whose Weyl group controlled the uncompactified billiard. The dynamics of the compactified billiard is described by the Weyl group $\bar{\mf{W}}$ of $\mf{g}(\bar{A})$, and the billiard region $\bar{\mc{B}}$ coincides with the fundamental domain $\bar{\mc{F}}$ of $\bar{\mf{W}}$. Thus, if $\mf{g}(\bar{A})$ is hyperbolic, then $\bar{\mc{B}}$ is of finite volume, yielding chaotic dynamics. If the Kac-Moody algebra $\mf{g}(\bar{A})$ is Lorentzian but not hyperbolic, then the billiard is non-chaotic. This is illustrated in Figure \ref{figure:WallCones}.

In many cases the billiard table is a simplex, but some dihedral angles are obtuse (positive inner product between two different walls) and the matrix $A_{ab}$ is not a proper Cartan matrix. It is however non-degenerate, so that it is possible to define a set of dominant ``coweights" $\Lambda^{A^{\p}\mu}$ such that
\beq
\omega_{A^{\p}\mu}\Lambda^{B^{\p}\mu} = {\delta^{B^{\p}}}_{A^{\p}},
\eeq
where $\omega$ is a dominant wall labelled by $A^{\p}$.  As in the standard Kac-Moody algebra case, these ``coweights" span the space of rays that lie within the wall cone, provided we only combine ``coweights" using non-negative coefficients.  A non-chaotic solution to the equations of motion, which corresponds to a null ray within the wall cone, exists if and only if there is at least one timelike and one spacelike ``coweight".  This condition is readily checked once the ``coweights" are in hand.  This technique was employed in \cite{Wesley:2006cd}.

When the billiard table is not a simplex and the number of walls is less than the dimension of $\mf{M}_{\be}$, then the theory is not chaotic, essentially because there are too few walls to prevent a ray from reaching infinity.

\begin{figure}[ht]
\begin{center}
\includegraphics[width=90mm]{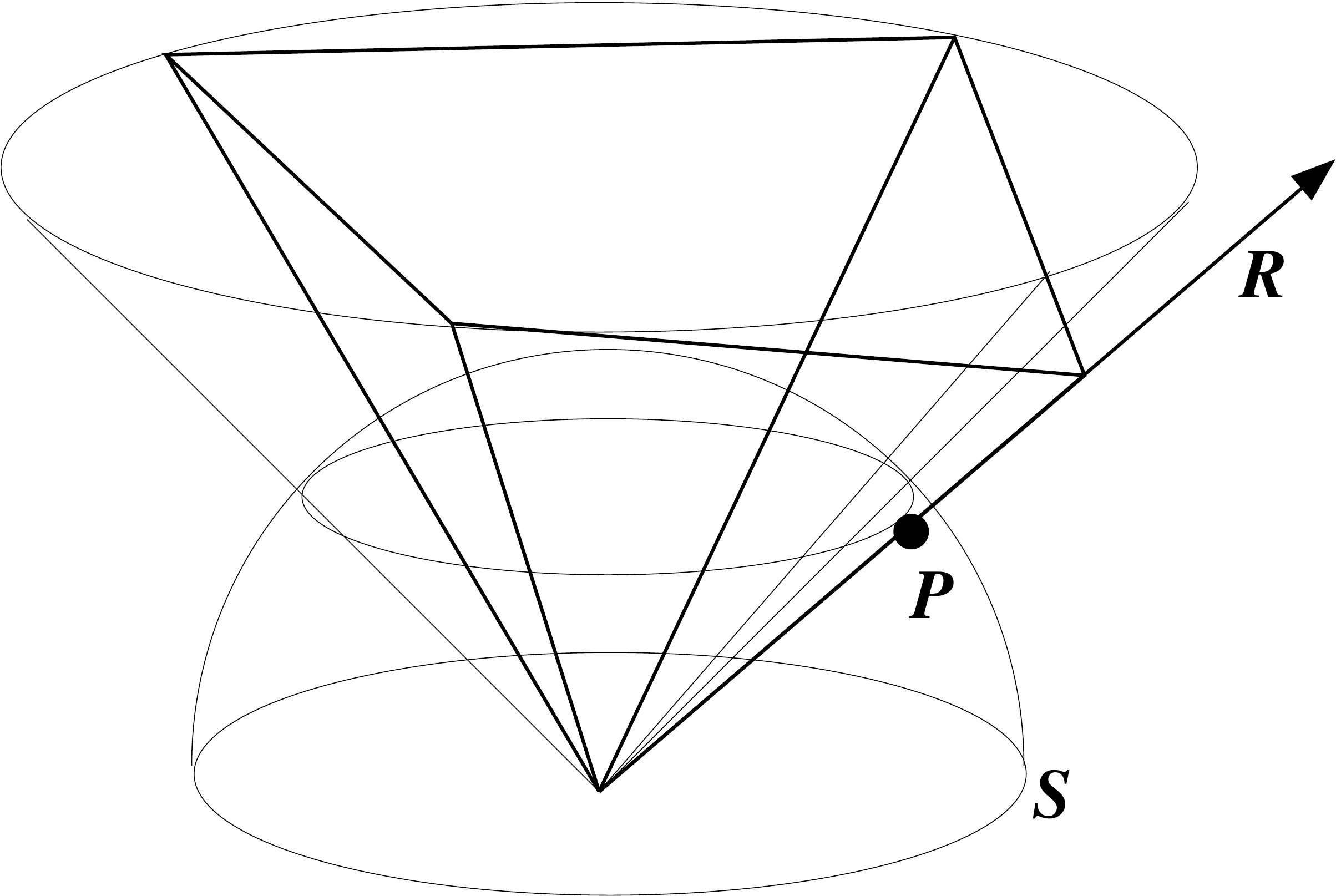}
\caption{The figure illustrates a non-simplex billiard table. To determine whether the theory is chaotic, it is sufficient to locate a spacelike ray. This is equivalent to maximising the Lorentzian norm on the unit sphere surrounding the origin.}
\label{figure:TooManyWalls}
\end{center}
\end{figure}

When the billiard table is not a simplex and the number of walls is greater than the dimension of $\mf{M}_{\be}$, the analysis becomes more complex.
This situation is illustrated in Figure \ref{figure:TooManyWalls}. One method to determine whether chaos is present is to successively remove dominant walls until the billiard region is again a simplex. If there is (at least) one way to do this such that the resulting structure is hyperbolic, then we can conclude that the full region is of finite volume, since reinserting the walls that were removed can never render the volume infinite. In a small number of cases, all wall removals lead to non-chaotic billiards and one cannot conclude immediately whether or not the volume of the billiard table is finite.  Another method is then to search numerically for whether a spacelike direction in the wall cone exists. We do this by maximising the Lorentzian norm of points on the unit sphere that lie within the wall cone.  If the maximal norm is negative, then no spacelike direction exists and the system is chaotic.

Also, as we have described in Section \ref{section:chambers} and \ref{section:E7}, it is sometimes possible to easily compute the volume of the billiard region exactly by making use of certain properties of the Weyl group $\mf{W}[\mf{g}]$, associated with the uncompactified theory and construct the associated gallery.

As a byproduct of the ``coweight construction" mentioned above, we can easily prove the following useful fact:

\begin{itemize}
\item {\bf Fact 3:} \emph{Whenever the billiard is described by a direct product $\mf{B}_{\text{fin}}\times \mf{B}_{\text{hyp}}$ of a finite and a hyperbolic Coxeter group, then the dynamics is non-chaotic.}
\end{itemize}

\noi This follows by noting that the metric is a direct sum of the metric of $\mf{B}_{\textit{fin}}$ and that of $\mf{B}_{\textit{hyp}}$ so the coweights associated with each factor define orthogonal subspaces.  Since the coweights associated with the finite factor are spacelike, there will always exist at least one spacelike intersection in the region inside the dominant walls. A different intuitive way to see this is to recall that the volume of the fundamental Weyl chamber of the first factor is finite after projection on the sphere and that of the second factor after projection on the hyperboloid. Two projections are needed to have a finite volume but there is only one here (on a hyperboloid living in the product space).

\section{Selected Examples}\label{section:Examples}

One of the main achievements of this paper is a complete list of billiard structures for general compactifications of all maximally-oxidised theories whose U-duality groups are split real forms.  This is presented in full in Appendix A, while in this section we present detailed calculations for a selection of these cases in an effort to illuminate the general results of the previous section.  We have chosen examples that highlight some particularly interesting or subtle issues in the analysis.

\subsection{The $F_4$-Sequence}
\label{section:F4}

For the theory with U-duality algebra $F_4$, the Kac-Moody algebraic structure is completely preserved by compactification. The oxidation endpoint of this theory is $D=6$ with field content given by two scalar fields, a dilaton $\phi$ and an axion $\chi$, two one-forms $A^{\pm}$, a two-form $B$ and, finally, a two-form $C$ with self-dual field strength $G=\star G$. The complete set of walls was found in \cite{Sophie} and is reproduced below.\footnote{We have corrected some minor misprints in \cite{Sophie}.} We have the following electric walls:
\beqa
{}e^{[B]}_{ij}(\be)  =  \be^{i}+\be^j-\phi, &\quad & e^{[C]}_{i}(\be) =  \be^{i}+\be^j,
\nn \\
{}e^{[\chi]}(\be) =  \phi, & \quad & e^{[A^{\pm}]}_{i}(\be)  =  \be^{i}\pm \f{1}{2}\phi,
\eqnlab{F4electricwalls}
\eqa
and magnetic walls:
\beqa
{}m^{[B]}_{ij}(\be)  =  \be^{i}+\be^{j}+\phi, &\quad  & m^{[C]}_{ij}(\be) =  \be^{i}+\be^{j},
\nn \\
{}m^{[\chi]}_{ijkl}(\be)  =  \be^{i}+\be^{j}+\be^{k}+\be^{l}-\phi, &\quad  & m^{[A^{\pm}]}_{ijk}(\be)  =  \be^{i}+\be^{j}+\be^{k}\mp \f{1}{2}\phi.
\eqnlab{F4magneticwalls}
\eqa
The subset of dominant walls is
\beqa
{}\al_{-1}(\be) \equiv  s_{54}(\be)=\be^5-\be^4,&\quad & \al_{0}(\be)\equiv  s_{43}(\be)=\be^4-\be^3,
\nn \\
{}\al_{1}(\be)\equiv  s_{32}(\be)=\be^3-\be^2,&\quad & \al_{2}(\be) \equiv  s_{21}(\be)=\be^2-\be^1,
\nn \\
{}  \al_{3}(\be) \equiv  e^{[A^-]}_1(\be)=\be^1-\f{1}{2}\phi,&\quad & \al_{4}(\be)\equiv  e^{[\chi]}(\be)=\phi.
\eqnlab{F4dominantwalls}
\eqa
These correspond to the simple roots of $F_4^{++}$ whose Dynkin diagram is displayed in Table \ref{table:F4++}. Thus, in the BKL-limit the billiard dynamics is controlled by the Weyl group of $F_4^{++}$ and since this is a hyperbolic Coxeter group, the theory exhibits chaotic behaviour.

We now analyze the billiard structure in the low-energy effective theory resulting from compactification on an arbitrary five-dimensional manifold $\mc{M}_5$. Since we only have $1$- and $2$-forms in the Lagrangian, only the first, second and third Betti numbers will affect the zero-mode spectrum after compactification.

\subsubsection{The $b_1(\mc{M}_5)=0$ Compactification}

Compactification on a manifold $\mc{M}_5$ with vanishing first Betti number $b_1(\mc{M}_5)=0$, projects out the dominant electric wall $e^{[A^-]}_1(\be)$. This pushes the electric wall $e^{[B]}_{12}(\be)$ to become dominant, and hence the third simple root $\al_{3}(\be)$ is replaced by
\beq
\tilde{\al}_3(\be)=\be^1+\be^2-\phi.
\eeq
This root is long, $(\tilde{\al}_{3}|\tilde{\al}_3)=2$, and has non-vanishing negative scalar products with $\al_1(\be)$ and $\al_{4}(\be)$,
\beq
(\tilde{\al}_3|\al_1)=(\tilde{\al}_3|\al_4)=-1.
\eeq
Hence, one can interpret it as a simple root of a new algebra. Since the fourth simple root is short, $(\al_4|\al_4)=1$, the resulting algebra is not simply laced, with the asymmetric part of the Cartan matrix given by
\beq
A_{34}=\f{2(\tilde{\al}_3|\al_4)}{(\tilde{\al}_3|\tilde{\al}_3)}=-1, \qquad A_{43}=\f{2(\al_4|\tilde{\al}_3)}{(\al_4|\al_4)}=-2.
\eeq
The new set of simple roots gives rise to the Dynkin diagram of the hyperbolic Kac-Moody algebra $B_4^{++}$, displayed in Table \ref{table:F4++}. We may therefore conclude that the dynamics in the BKL-limit (in the regime of intermediate asymptotic) remains chaotic after compactification. The simple roots $\{\al_{-1},\al_0,\al_{1},\al_2,\tilde{\al}_3,\al_4\}$ all correspond to real positive roots of $F_4^{++}$ and therefore this indeed corresponds to a regular embedding of $B_4^{++}$ into $F_4^{++}$,
\beq
\bar{\mf{g}}[b_1]=B_4^{++}\hs \subset \hs F_4^{++}.
\eeq
This example thus exhibits a ``jump'' between two oxidation chains. After compactification, the billiard structure is that of $B_4^{++}$, which usually is associated with the billiard of a six-dimensional theory with $p$-form spectrum given by a dilaton, a Maxwell field and a 2-form. These two, seemingly different, theories exhibit identical behaviour in the BKL-limit if the $F_4$-theory is compactified on $\mc{M}_5$ with $b_1(\mc{M}_5)=0$. It is tempting to speculate that the preserved algebraic structure reflects that this particular reduction is a consistent truncation of the original theory.

\subsubsection{The $b_1(\mc{M}_5)=b_2(\mc{M}_5)=0$ Compactification}

We proceed with the analysis of the $F_4$-sequence to reduction on a manifold with $b_1(\mc{M}_5)=b_2(\mc{M}_5)=0$. The original set of dominant walls is unchanged by taking $\mc{M}_5$ with $b_2(\mc{M}_5)=0, b_3(\mc{M}_5)=0$ or $b_2(\mc{M}_5)=b_3(\mc{M}_5)=0$, so these constraints have no influence on the chaotic behaviour. However, the combination of $b_1(\mc{M}_5)=0$ and $b_2(\mc{M}_5)=0$ drastically changes the set of non-gravitational walls of the theory and so is sufficient to remove chaos.

As we saw in the first case above, the constraint $b_{1}(\mc{M}_5)=0$ removes the dominant wall $e^{[A^{-}]}_1(\be)$ and replaces it with the electric wall $e^{[B]}_{12}(\be)$, resulting in the hyperbolic algebra $B_4^{++}$. Now also the electric wall $e^{[B]}_{12}(\be)$ is removed by the additional constraint $b_2(\mc{M}_5)=0$. Actually, we also kill the magnetic walls of the axion $\chi$ and of the one-forms $A^{\pm}$ so it turns out that the new dominant wall is the gravity wall $G_{145}(\be)$, i.e.,
\beq
\tilde{\al}_3(\be)=2\be^1+\be^2+\be^3.
\eeq
This positive root can be interpreted as a simple root of a new algebra, and connects with a single link to $\al_0$ and $\al_2$, resulting in the Dynkin diagram of $A_3^{++}$ which is hyperbolic. However, we also have the axion electric wall $e^{[\chi]}(\phi)$ which has vanishing scalar product with all the other simple roots and hence represents a disconnected $A_1$-factor. The resulting algebra is therefore the direct sum
\beq
\bar{\mf{g}}[b_1,b_2]=A_3^{++}\oplus A_1 \subset F_4^{++}.
\eeq
Because of {\bf Fact 3} (Section \ref{section:ChaoticProperties}) this theory is non-chaotic.

\subsection{The $E_7$-Sequence}
\label{section:E7}

The compactifications of the $E_7$-theory considered here provide a complete counterpart to the examples studied in the previous section since no compactification yields a Coxeter simplex\footnote{Except those compactifications which do not change the dominant walls at all.}.  Here we show explicitly that for these compactifications the dominant walls of the massless $p$-form spectrum violate  the conditions for a valid root system of a Kac-Moody algebra. We study in detail the underlying Coxeter group and determine the structure of the billiard table. We find that the billiard region after compactification is not a fundamental domain, but corresponds to a gallery of finite length inside the Cartan subalgebra of $E_7^{++}$. We also show that for both compactifications the billiard group $\mf{B}$ corresponds to a non-standard presentation of $E_7^{++}$.

\subsubsection{Dominant Wall Structure After Compactification}
\label{section:dominantwallsE7}

The $E_7$-sigma model oxidises to a non-supersymmetric consistent truncation of maximal $D=9$ supergravity with bosonic field content given by a dilaton $\phi$, a Maxwell field $A$ and a 3-form $C$. The electric and magnetic walls are \cite{Sophie}:
\beqa
e_i^{[A]}(\be) = \be^{i}-\f{2\sqrt{2}}{\sqrt{7}}\phi,&\quad & e_{ijk}^{[C]}(\be)= \be^{i}+\be^{j}+\be^{k}+\f{\sqrt{2}}{\sqrt{7}}\phi,
\nn \\
{}m_{i_1,\dots,i_6}^{[A]}(\be)= \be^{i_1}+\cdots + \be^{i_6}+\f{2\sqrt{2}}{\sqrt{7}}\phi, & \quad & m_{ijkl}^{[C]}(\be)= \be^{i}+\be^{j}+\be^{k}+\be^{l}-\f{\sqrt{2}}{\sqrt{7}}\phi.
\nn \\
\eqnlab{E7electricwalls}
\eqa
The dominant walls are
\beqa
{}\al_{-1}(\be)= s_{87}(\be)=\be^8-\be^7, &\quad & \al_{0}(\be) = s_{76}(\be)=\be^7-\be^6,
\nn \\
{} \vdots & \quad  & \vdots
\nn \\
{} \al_4(\be)=s_{32}(\be)=\be^{3}-\be^{2},&\quad & \al_{5}(\be)= s_{21}(\be)=\be^2-\be^1,
\nn \\
{}\al_6(\be)= e_1^{[A]}(\be)=\be^1-\f{2\sqrt{2}}{\sqrt{7}}\phi,&\quad & \al_7(\be)= e_{123}^{[C]}(\be)=\be^1+\be^2+\be^3+\f{\sqrt{2}}{\sqrt{7}}\phi.
\nn \\
\eqnlab{E7dominantwalls}
\eqa
The Dynkin diagram constructed from these simple roots coincides with the one of $E_7^{++}$ and is shown in Table \ref{table:E7++1}.

\vspace{.3cm}

\noi {\bf \small{The $b_1(\mc{M}_8)=0$ Compactification}}

\vspace{.2cm}

\noi Taking an eight-dimensional internal manifold $\mc{M}_8$ with vanishing first Betti number  removes the dominant electric wall $e_1^{[A]}(\be)$ and the new dominant wall is the magnetic wall of the 3-form
\beq
\tilde{\al}_6(\be)=m^{[C]}_{1234}(\be)=\be^1+\be^2+\be^3+\be^4-\f{\sqrt{2}}{\sqrt{7}}\phi.
\eeq
However, this positive root is not acceptable as a simple root of a Kac-Moody subalgebra of $E_7^{++}$ since its scalar product with $\al_7$ is positive,
\beq
(\tilde{\al}_6|\al_7)=1.
\eeq
The billiard table has one obtuse dihedral angle.
Although the dominant wall set does not constitute a set of simple roots of a hyperbolic Kac-Moody algebra we may still compute the associated dominant ``coweights'' (as described in Section \ref{section:ChaoticProperties}), and we find that they are all timelike, revealing that the resulting dynamics is in fact chaotic.

\vspace{.3cm}

\noi {\bf \small{The $b_3(\mc{M}_8)=0$ Compactification}}

\vspace{.2cm}

\noi The case $b_2(\mc{M}_8)=0$ has no effect on the original wall system so we proceed directly to $b_3(\mc{M}_8)=0$. The dominant electric wall $e_{123}^{[C]}(\be)$ is projected out which leaves us with three possible candidates for the new dominant wall
\beqa
{} m_{123456}^{[A]}(\be)&=& \be^1+\cdots +\be^6 +\f{2\sqrt{2}}{\sqrt{7}}\phi
\nn \\
{} m_{1234}^{[C]}(\be)&=& \be^1+\cdots + \be^4 -\f{\sqrt{2}}{\sqrt{7}}\phi
\nn \\
{}G_{178}(\be)&=& 2\be^1+\be^2 +\cdots + \be^6.
\eqa
The gravity wall $G_{178}$ is subdominant because it can be obtained as the linear combination
\beq
G_{178}(\be)=m_{123456}^{[A]}(\be)+e_1^{[A]}(\be).
\eeq
For the remaining two walls it turns out that none can be obtained as an integral linear combination in terms of the other walls. Hence, both walls $m_{123456}^{[A]}$ and $m_{1234}^{[C]}$ are actually dominant. The full set of dominant walls therefore does not define a simplex.  Furthermore, there is one obtuse dihedral angle.  One cannot interpret the walls as simple roots of a valid root system.

The resulting structure is best described by choosing $m_{123456}^{[A]}$ as a new simple root $\tilde{\al}_7$ and leaving the remaining magnetic wall $m_{1234}^{[C]}$ outside of the algebraic structure. Thus we take
\beq
\tilde{\al}_7(\be)=\be^1+\cdots +\be^6+\f{2\sqrt{2}}{\sqrt{7}}\phi.
\eeq
This is a long root, $(\tilde{\al}_7|\tilde{\al}_7)=2$, and has non-vanishing negative scalar products with $\al_0$ and $\al_6$,
\beq
(\tilde{\al}_7|\al_0)=-1,\qquad \qquad (\tilde{\al}_7|\al_6)=-1,
\eeq
implying that the new node in the Dynkin diagram is connected with a single link to nodes $0$ and $6$. This therefore gives the Dynkin diagram of $A_7^{++}$ which is hyperbolic. Because of the remaining magnetic wall $m_{1234}^{[C]}$ the complete set of dominant walls defines a nine-dimensional bounded region, which is smaller than the fundamental domain of $A_7^{++}$, and hence has also finite volume: adding the extra wall $m_{1234}^{[C]}$ can never render the volume of the billiard infinite and therefore the dynamics is still chaotic.

The algebra $A_7^{++}$ is usually associated with the billiard of pure gravity in $D=10$ dimensions, so it is interesting that we here find an embedding of $A_7^{++}$ into $E_7^{++}$ which involves both a magnetic and an electric simple root (although it does not describe the complete dominant wall structure for this theory).

\subsubsection{Non-Standard Presentations of $E_7^{++}$}

For the compactifications studied in the previous section, there is no longer an interpretation of the wall structure in the reduced theory as simple roots of a Kac-Moody algebra. The billiard reflections of the compactified theory, however, still generate the Coxeter group $E_7^{++}$, although they do not provide a standard presentation.

Let $s_i$ $(i=-1, 0, 1, \dots, 7)$ denote the nine fundamental reflections of the Weyl group $E_7^{++}$.\footnote{As is common in the literature we refer to the crystallographic Coxeter group by the same name as for the associated Kac-Moody algebra.} As explained in Section \ref{section:BKL}, this group is completely determined by its Coxeter exponents $m_{ij}$, and the defining relations $(s_is_j)^{m_{ij}}=1$. This gives the following Coxeter presentation of $E_7^{++}$:
\beq
E_7^{++}=\big< s_{-1}, s_0, s_1, \dots, s_7\ \big|\ (s_i s_j)^{m_{ij}}=1,\ i, j=-1, 0, 1, \dots, 7\ \big>,
\eeq
where the Coxeter exponents $m_{ij}$ follow from the dihedral angles of the fundamental Weyl chamber of $E_7^{++}$ and can be read off from its Coxeter graph/Dynkin diagram, displayed in Table \ref{table:E7++1}.

\vspace{.3cm}

\noi {\bf \small{The $b_1(\mc{M}_8)=0$ Compactification}}

\vspace{.2cm}

\begin{figure}[ht]
\begin{center}
\includegraphics[width=90mm]{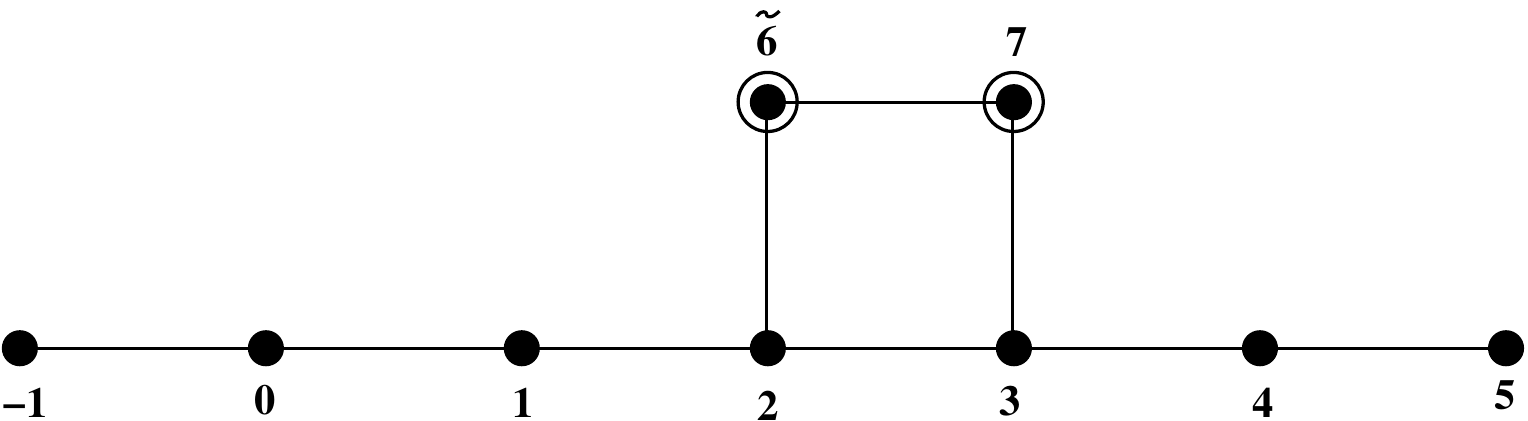}
\caption{The Coxeter graph of the formal Coxeter group $\mf{C}$ generated by the fundamental reflections $s_{-1}, s_0, s_1, \dots, \tilde{s}_6, s_7$ of the billiard associated with $b_1(\mc{M}_8)=0$. The circles around the nodes $\tilde{6}$ and $7$ indicate that the dihedral angle between the associated walls is obtuse. The billiard group $\mf{B}$ is a quotient of $\mf{C}$ and turns out to be $E_7^{++}$.}
\label{figure:E7++b1}
\end{center}
\end{figure}

\noi Consider now compactification on a manifold $\mc{M}_8$ with $b_1(\mc{M}_8)=0$. After compactification, the dominant set of wall forms is $\{\al_{-1}, \al_0, \al_1, \dots, \tilde{\al}_6, \al_7\}$. The new dominant walls give rise to a set of reflections
\beq
\mc{S}=\{s_{-1}, s_0, s_1, \dots, \tilde{s}_6, s_7\}.
\eeq
We wish to find which is the Coxeter group generated by these reflections, i.e., the corresponding billiard group $\mf{B}$. The formal Coxeter group $\mf{C}$ associated with these reflections is determined by the Coxeter graph in Figure \ref{figure:E7++b1} as the quotient group  $\mf{C}=\tilde{\mf{C}}/\mf{N}$, where $\mf{N}$ is the normal subgroup generated by the relations among the elements of the freely generated group $\tilde{\mf{C}}$ as encoded in the Coxeter exponents defined by the Coxeter graph.

This is not the end of the story, though, because there are additional relations between the generators, apart from the standard Coxeter relations. Indeed, the new fundamental reflection $\tilde{s}_6$ can be written in terms of the original reflections as follows
\beq
\tilde{s}_6=s_7 s_3 s_4 s_5 s_6 s_5 s_4 s_3 s_7,
\eqnlab{stilde6}
\eeq
which in turn can be inverted to obtain the old reflection $s_6$ in terms of the new ones,
\beq
s_6 = s_5 s_4 s_3 s_7 \tilde{s}_6 s_7 s_3 s_4 s_5.
\eqnlab{s6}
\eeq
This implies that there is one (independent) extra relation among the generators of the billiard group $\mf{B}$
\beq
\big(s_5 s_4 s_3 s_7 \tilde{s}_6 s_7 s_3 s_4 s_5 s_2)^{2}=1.
\eqnlab{b1relation}
\eeq
This relation is inherited, via \Eqnref{s6}, from the relation $(s_6 s_2)^{2}=1$ in the original Coxeter group $E_7^{++}$. Consider the normal subgroup $\mf{I}$ of  $\mf{C}$ generated by
\beq
\big(s_5 s_4 s_3 s_7 \tilde{s}_6 s_7 s_3 s_4 s_5 s_2)^{2}.
\eeq
Taking the quotient by this subgroup gives a group which is isomorphic to $E_7^{++}$,
\beq
E_7^{++}=\mf{C} \big/  \mf{I}
\eeq
since it contains all the fundamental reflections $s_i$ $(i=-1, 0, 1, \dots, 7)$ and their relations.  This is, however, a non-standard representation of the Coxeter group $E_7^{++}$.

We see therefore that after compactification the billiard dynamics is still controlled by the hyperbolic Coxeter group $E_7^{++}$, albeit with a non-standard presentation. This indicates that the Coxeter group structure is a more rigid structure which may survive even when the billiard table is not the original Coxeter polyhedron.

\vspace{.3cm}

\noi {\bf \small{The $b_3(\mc{M}_8)=0$ Compactification}}

\vspace{.2cm}

\begin{figure}[ht]
\begin{center}
\includegraphics[width=90mm]{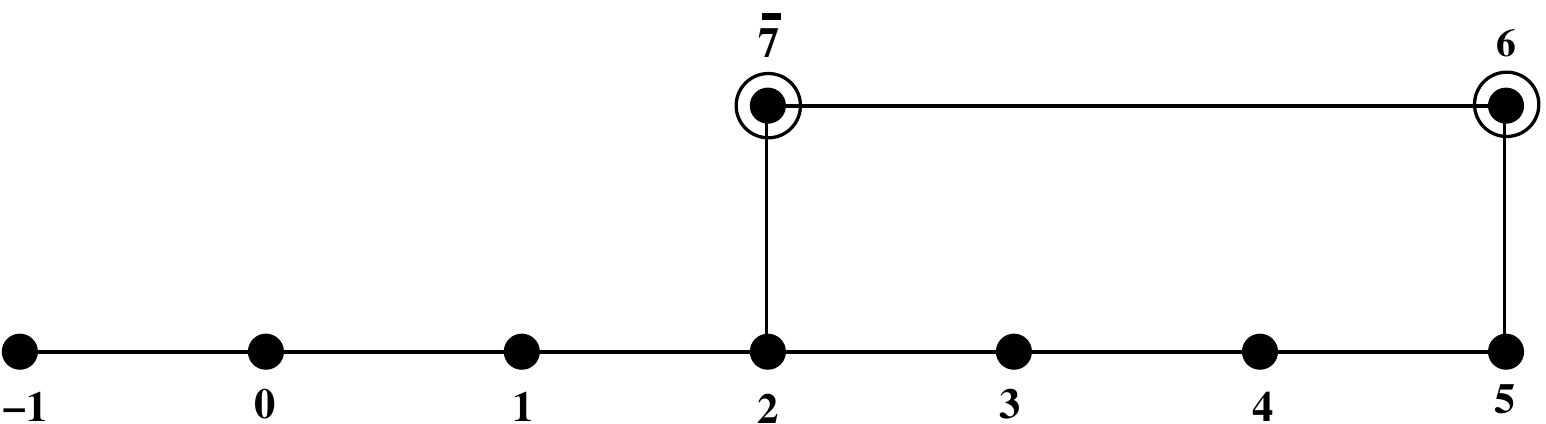}
\caption{The Coxeter graph of the formal Coxeter group $\mf{R}$ generated by the fundamental reflections $s_{-1}, s_0, s_1, \dots, s_6, \bar{s}_7$ of the billiard associated with $b_3(\mc{M}_8)=0$ (leaving away the magnetic wall $m_{123456}^{[A]}$). The circles around the nodes $\bar{7}$ and $6$ indicate that the dihedral angle between the associated walls is obtuse. The billiard group $\mf{B}$ is a quotient of $\mf{R}$ and turns out to be $E_7^{++}$.}
\label{figure:E7++b3}
\end{center}
\end{figure}

\noi We now determine the Coxeter group of the $b_3(\mc{M})=0$ compactification. It is convenient to initially leave the magnetic wall $m_{123456}^{[A]}$ outside of the analysis. Thus, we take
\beq
\bar{\al}_7=m_{1234}^{[C]}=\be^{1}+\cdots +\be^4-\f{\sqrt{2}}{\sqrt{7}}\phi,
\eeq
as the new ``simple root''. We denote the Coxeter group freely generated by the set $\mc{R}=\{s_{-1}, s_0, s_1, \dots, s_6, \bar{s}_7\}$ by $\tilde{\mf{R}}$,  the normal subgroup defined by the Coxeter exponents corresponding to the dihedral angles of the billiard table by $\mf{P}$, and the corresponding formal Coxeter group by $\mf{R}=\tilde{\mf{R}}/\mf{P}$. The associated Coxeter graph is displayed in Figure \ref{figure:E7++b3}.

Similarly to the previous case we can express the original fundamental reflection $s_7$ in terms of the new ones as
\beq
s_7=s_3s_4s_5s_6 \bar{s}_7 s_6 s_5s_4 s_3.
\eeq
The relation $(s_7 s_2)^{2}=1$ then induces the following relation among the generators of $\mf{R}$ :
\beq
\big(s_3s_4s_5s_6 \bar{s}_7 s_6 s_5s_4 s_3 s_2\big)^2=1,
\eeq
which is not a standard Coxeter-type relation. Again, this implies that $\mf{R}$ has a normal subgroup $\mf{J}$, generated by
\beq
\big(s_3s_4s_5s_6 \bar{s}_7 s_6 s_5 s_4 s_3 s_2\big)^2,
\eeq
and hence by taking the quotient by $\mf{J}$ we find yet another non-standard presentation of $E_7^{++}$,
\beq
E_7^{++}=\mf{R}\big/ \mf{J}.
\eeq
Adding the reflection in the wall that has been dropped does not modify the result since this reflection can in fact be expressed in terms of the other ones. Thus, this compactification also preserves the Coxeter group $E_7^{++}$ of the billiard, even though it yields a system of walls that cannot be interpreted as the simple roots of $E_7^{++}$.

\subsubsection{The Volume of the Billiard Gallery in the Case $b_1(\mc{M}_8)=0$}

Here we give an explicit description of the billiard region as a union of images of the fundamental region under the Coxeter group of the noncompact theory, for the compactification with $b_1(\mc{M}_8)=0$ studied above.  We compute the volume of the billiard gallery and show that it is finite, which implies that the reduced theory is chaotic.

As usual, we associate a fundamental reflection $s_i\in E_7^{++}$ to each of the simple roots of $E_7^{++}$. As we have seen, the compactification with $b_1(\mc{M}_8)=0$ amounts to dropping the wall located at $\al_6(\be)=0$ and replacing it by $\tilde{\al}_6(\be)=0$.  Thus, one allows for the possibility that the billiard ball moves on the negative side of the electric wall $\al_6(\be)=0$, i.e., to the region $\al_6(\be)\leq 0$. One can go to that region from the fundamental Weyl chamber $\mc{F}$ of $E_7^{++}$ by performing a Weyl reflection in the electric wall which is removed. Starting from the region $\mc{B}_1\equiv \mc{F}$, we shall thus apply the reflection $s_6$ to all the bounding walls of $\mc{B}_1$. The result is
\beqa
{}s_6(\al_6)=-\al_6\quad &\Longleftrightarrow &\quad \al_6(\be)\leq 0,
\nn \\
{}s_6(\al_5)=\al_5+\al_6 \quad & \Longleftrightarrow & \quad (\al_5+\al_6)(\be)\geq 0,
\eqa
with all other walls left invariant by $s_6$. In this way we obtain a new region $\mc{B}_2$, defined by the image of $s_6$ on $\mc{F}$, i.e.,
\beqa
{}\mc{B}_2 & =& s_6\cdot \mc{F}
\nn \\
{}& = & \big\{\be\in\mf{h}\ \big|\ \al_{-1}(\be)\geq 0, \dots, \al_4(\be)\geq 0, \al_7(\be)\geq 0, (\al_5+\al_6)(\be)\geq 0, \al_6(\be)\leq 0 \big\}.
\nn \\
\eqa
At this point we still have $\mc{B}_1\cup \mc{B}_2\subset \mc{B}$ as a proper inclusion, i.e., we have not yet covered all of $\mc{B}$. To proceed we must also allow for the possibility that $(\al_5+\al_6)(\be)\leq 0$, which corresponds to performing a Weyl reflection in the wall $(\al_5+\al_6)(\be)=0$. A quick calculation reveals that the relevant reflection is the combination $s_5s_6s_5\in E_7^{++}$, as is evident from
\beq
s_5 s_6 s_5 (\al_5+\al_6)=-(\al_5+\al_6).
\eeq
The remaining reflections which are affected by $s_5s_6s_5$ are
\beqa
{}s_5s_6s_5 (\al_4) = \al_4+\al_5+\al_6\quad &\Longleftrightarrow&\quad (\al_4+\al_5+\al_6)(\be)\geq 0,
\nn \\
{}s_5s_6s_5(\al_6)=-\al_5 \quad &\Longleftrightarrow & \quad \al_5(\be)\geq 0.
\eqa
We see that there is a new constraint $(\al_4+\al_5+\al_6)(\be)\geq 0$, and we obtain the region $\mc{B}_3$, defined as
\beqa
{}\mc{B}_3 &=& s_5s_6s_5\cdot \mc{B}_2
\nn \\
{}&=& \big\{\be\in\mf{h}\ \big|\ \al_{-1}(\be)\geq 0, \dots, \al_3(\be)\geq 0, \al_5(\be)\geq 0, \al_7(\be)\geq 0,
\nn \\
{}&  & (\al_5+\al_6)(\be)\leq 0, (\al_4+\al_5+\al_6)(\be)\geq 0\big\}.
\nn \\
\eqa
The next step is to allow for the possibility $(\al_4+\al_5+\al_6)(\be)\leq 0$, which corresponds to a Weyl reflection with respect to the wall $(\al_4+\al_5+\al_6)(\be)=0$. The particular combination of fundamental reflections corresponding to this reflection is $s_4s_5s_6s_5s_4\in E_7^{++}$,
\beq
s_4s_5s_6s_5s_4(\al_4+\al_5+\al_6)=-(\al_4+\al_5+\al_6).
\eeq
The action of $s_4s_5s_6s_5s_4$ on the remaining (non-invariant) walls of $\mc{B}_3$ yields
\beqa
{}s_4s_5s_6s_5s_4(\al_3)=\al_3+\al_4+\al_5+\al_6\quad & \Longleftrightarrow &\quad (\al_3+\al_4+\al_5+\al_6)(\be)\geq 0,
\nn \\
{}s_4s_5s_6s_5s_4(\al_5)=\al_5 \quad & \Longleftrightarrow &\quad \al_5(\be)\geq 0,
\nn \\
{}s_4s_5s_6s_5s_4(\al_5+\al_6)=-\al_4 \quad &\Longleftrightarrow &\quad \al_4(\be)\geq 0.
\eqa
As before, we see that a new constraint $(\al_3+\al_4+\al_5+\al_6)(\be)\geq 0$ arises. Putting things together we find that the region $\mc{B}_4$ is given by
\beqa
\mc{B}_4 &=& s_4s_5s_6s_5s_4\cdot \mc{B}_3
\nn \\
{}&=& \big\{\be\in\mf{h}\ \big| \ \al_{-1}(\be)\geq 0, \dots, \al_2(\be)\geq 0, \al_4(\be)\geq 0, \al_5(\be)\geq 0, \al_7(\be)\geq 0,
\nn \\
{}& &  (\al_4+\al_5+\al_6)(\be)\leq 0, (\al_3+\al_4+\al_5+\al_6)(\be)\geq 0 \big\}.
\nn \\
\eqa
Following the same route, we now allow for $(\al_3+\al_4+\al_5+\al_6)(\be)\leq 0$. The associated Weyl reflection is $\bar{s}\equiv s_3s_4s_5s_6s_5s_4s_3$, which follows from
\beq
\bar{s}(\al_3+\al_4+\al_5+\al_6)=-(\al_3+\al_4+\al_5+\al_6).
\eeq
The action of $\bar{s}$ on $\mc{B}_4$ then yields
\beqa
{}\bar{s}(\al_2)=\al_2+\al_3+\al_4+\al_5+\al_6  & \Longleftrightarrow &  (\al_2+\al_3+\al_4+\al_5+\al_6)(\be)\geq 0,
\nn \\
{}\bar{s}(\al_4)=\al_4  & \Longleftrightarrow &  \al_4(\be)\geq 0,
\nn \\
{}\bar{s}(\al_5)=\al_5  &\Longleftrightarrow &   \al_5(\be)\geq 0,
\nn \\
{}\bar{s}(\al_7)=\al_3+\al_4+\al_5+\al_6+\al_7 & \Longleftrightarrow& (\al_3+\al_4+\al_5+\al_6+\al_7)(\be)\geq 0,
\nn \\
{}\bar{s}(\al_4+\al_5+\al_6)=-\al_3 &\Longleftrightarrow &  \al_3(\be)\geq 0.
\nn \\
\eqa
Here we see that the new constraint $(\al_3+\al_4+\al_5+\al_6+\al_7)(\be)\geq 0$, replacing $\al_7(\be)\geq 0$, coincides with the constraint $\tilde{\al}_6(\be)\geq 0$, associated with the new ``simple root'' $\tilde{\al}_6= m^{[C]}_{1234}(\be)$, and we may thus conclude that we have reached the far end of the region $\mc{B}$. However, we also uncover a new wall inequality $(\al_2+\al_3+\al_4+\al_5+\al_6)(\be)\geq 0$, arising because of the ``vertex'' adjoining nodes $2, 4$ and $7$ to node $3$ in the Dynkin diagram of $E_7^{++}$ (see Table \ref{table:E7++1}). The region $\mc{B}_5$ is thus given by
\beqa
{}\mc{B}_5&=& \bar{s}\cdot \mc{B}_4
\nn \\
{}&=& \big\{\be\in\mf{h}\ \big| \ \al_{-1}(\be)\geq 0, \al_0(\be)\geq 0, \al_1(\be)\geq 0, \al_3(\be)\geq 0, \al_4(\be)\geq 0,
\nn \\
{}& & \al_5(\be)\geq 0,(\al_3+\al_4+\al_5+\al_6)(\be)\leq 0, (\al_2+\al_3+\al_4+\al_5+\al_6)(\be)\geq 0,
\nn \\
{} & & (\al_3+\al_4+\al_5+\al_6+\al_7)(\be)\geq 0\big\}.
\nn \\
\eqa
Although we have reached one edge of the region $\mc{B}$, we have not yet exhausted it. To proceed, we must reflect in the wall $(\al_2+\al_3+\al_4+\al_5+\al_6)(\be)=0$, for which the desired reflection is $s_2s_3s_4s_5s_6s_5s_4s_3s_2$. We shall not display any more details of the calculations but merely state the result. We obtain a new region $\mc{B}_6$, given by
\beqa
{} \mc{B}_6 &=& s_2s_3s_4s_5s_6s_5s_4s_3s_2 \cdot \mc{B}_5
\nn \\
{}& = & \big\{\be\in \mf{h}\ \big|\ \al_{-1}(\be)\geq 0, \al_0(\be)\geq 0, \al_2(\be)\geq 0, \dots, \al_5(\be)\geq 0,
\nn \\
{}& & (\al_2+\al_3+\al_4+\al_5+\al_6)(\be)\leq 0, (\al_1+\al_2+\al_3+\al_4+\al_5+\al_6)(\be)\geq 0,
\nn \\
{}& & (\al_3+\al_4+\al_5+\al_6+\al_7)(\be)\geq 0\big\}.
\nn \\
\eqa
It is satisfactory to see that the far edge $(\al_3+\al_4+\al_5+\al_6+\al_7)(\be)\geq 0$ of the region $\mc{B}$ is preserved. We proceed to reflect in $(\al_1+\al_2+\al_3+\al_4+\al_5+\al_6)(\be)=0$, which yields
\beqa
{}\mc{B}_7 &=& s_1s_2s_3s_4s_5s_6s_5s_4s_3s_2s_1\cdot \mc{B}_6
\nn \\
{}&=& \big\{\be\in\mf{h} \ \big| \ \al_{-1}(\be)\geq 0, \al_1(\be)\geq 0, \dots, \al_5(\be)\geq 0, (\al_1+\cdots +\al_6)(\be)\leq 0,
\nn \\
{}& & (\al_0+\cdots +\al_6)(\be)\geq 0, (\al_3+\al_4+\al_5+\al_6+\al_7)(\be)\geq 0\big\}.
\nn \\
\eqa
Proceeding in this fashion we find the two additional regions
\beqa
{}\mc{B}_8 &=& s_0s_1s_2s_3s_4s_5s_6s_5s_4s_3s_2s_1s_0\cdot \mc{B}_7
\nn \\
{}& =&  \big\{\be\in\mf{h}\ \big| \ \al_0(\be)\geq 0,\dots , \al_5(\be)\geq 0, (\al_0+\cdots +\al_6)(\be)\leq 0,
\nn \\
{} & & (\al_{-1}+\cdots +\al_6)(\be)\geq 0, (\al_3+\al_4+\al_5+\al_6+\al_7)(\be)\geq 0 \big\},
\nn \\
\eqa
and
\beqa
{}\mc{B}_9 &=& s_{-1}s_0s_1s_2s_3s_4s_5s_6s_5s_4s_3s_2s_1s_0s_{-1}\cdot \mc{B}_8
\nn \\
{}& =&  \big\{\be\in\mf{h}\ \big| \ \al_{-1}(\be)\geq 0,\dots , \al_5(\be)\geq 0, (\al_{-1}+\cdots +\al_6)(\be)\leq 0,
\nn \\
{} & & (\al_3+\al_4+\al_5+\al_6+\al_7)(\be)\geq 0 \big\}.
\nn \\
\eqa
We see now that there is no longer any remaining wall to reflect in without ending up outside of $\mc{B}$. Thereby we have achieved our goal in covering all of $\mc{B}$ by a finite sequence of chambers, and we have the union
\beq
\mc{B}=\bigcup_{i=1}^{9}\mc{B}_i.
\eeq
Since any two consecutive chambers $\mc{B}_i$ and $\mc{B}_{i+1}$ are adjacent, the new billiard region is naturally described by a gallery $\Gamma$ connecting $\mc{B}_1$ with $\mc{B}_9$,
\beq
\Gamma\hs :\hs \mc{B}_1, \mc{B}_2, \dots, \mc{B}_8, \mc{B}_9.
\eeq
The length of $\Gamma$ is $k=9$ so the volume of $\mc{B}$ is
$9\cdot \text{vol}\ \mc{F}$,
which is finite.
\section{Concluding Remarks}
\label{section:Conclusions}
We have presented the complete compactification analysis for all of the Einstein-dilaton-$p$-form systems associated with split real forms of finite-dimensional Lie algebras. These are the maximal oxidations of models whose U-duality groups in three spacetime dimensions are finite simple Lie groups.  We have extended previous studies of the billiards associated with these models by including the effects of compactification on manifolds of nontrivial topology, which influences the billiard system (relevant to the pre-Planckian regime) through the zero-mode spectrum.

In most cases the algebraic interpretation of the dominant wall system as simple roots of a Kac-Moody algebra is unavailable, but in all cases the Coxeter group structure is preserved. This appears to indicate that the discrete symmetry groups revealed in the BKL-limit are more rigid than the continous Kac-Moody symmetries. It is possible that this fact is intimately related to the underlying quantum U-duality groups $\mc{U}(\mbb{Z})$ which indeed contain Coxeter groups as subgroups.

On the mathematical side we have found that compactification reveals interesting new structures of the cosmological billiard. In particular, we showed that the geometric reflections with respect to the dominant walls after compactification generate a Coxeter group which generically is described by a non-standard presentation, this being closely linked to the fact that the new billiard region is not a fundamental domain of the reflection group. Using technology from the theory of buildings we also found that this billiard region has a natural description as a gallery, i.e., an ordered sequence of Weyl chambers inside the Cartan subalgebra of the overextended U-duality algebra $\mf{u}^{++}$, whose Weyl group controlled the billiard dynamics before compactification.

One of the motivations for this work was to determine whether compactification on topologically nontrivial manifolds may eliminate chaos, which would be favorable for big crunch/big bang models which rely on smooth data during the collapse towards the singularity \cite{Khoury:2001wf,Steinhardt:2001st,Buonanno:1998bi,Gasperini:2002bn}. Our results indicate that this can indeed be achieved for some compactifications. For the $E_8$-sequence, which is relevant for M-theory and IIA/IIB string theory, chaos can be removed by taking an internal manifold $\mc{M}$ with $b_3(\mc{M})=b_4(\mc{M})=0$. This seems to imply that the simplest Calabi-Yau or $G_2$-holonomy manifolds are unsuitable for these constructions. For the heterotic string, whose billiard is described by $B_8^{++}$, chaos is removed for cases where all spatial dimensions are compact, with the exception of those for which only the first, or only the second, Betti number of the internal manifold vanishes. More generally, for models with no dilaton in dimensions $\leq 10$, chaos cannot be removed by the methods described here since the gravitational wall always remains and pure gravity is known to be chaotic \cite{Demaret:1986su}. However, for models with no dilaton in dimension $> 10$, or for models with dilatons, chaos can certainly be removed for example by dropping all the wall forms since the resulting models are non-chaotic \cite{Demaret:1986su,Demaret:1986ys,BelKal,Andersson:2000cv,Subcritical} (the axion wall that remains in the case of $F_4$ does not invalidate the conclusion).   We anticipate that these results will help to serve as a ``selection principle'' for future investigations of big crunch/big bang transitions in a compact setting.

Although we have focussed on theories exhibiting split U-duality symmetries in three dimensions, our analysis also applies to the case of non-split real forms. This is due to the fact that when the U-duality algebra $\mf{u}$ is non-split, and the restricted root system of $\mf{u}$ is reduced, the billiard dynamics is controlled by the overextension $\mf{f}^{++}$ of the maximal split ``subalgebra'' $\mf{f}$ of $\mf{u}$ \cite{MHBJ} (see also \cite{LivingReview}). All the algebras $\mf{f}^{++}$ are part of the classification considered here, and hence our results cover also these cases. The analysis can easily be extended also to the few cases where the restricted root system of $\mf{u}$ is non-reduced. 

Finally, we would like to mention that our work appears to be closely related to that of \cite{Ganor,Carlevaro}. These authors consider an M-theory setup where all spatial dimensions are compactified on certain $\mbb{Z}_n$-orbifolds of $T^{10}$, i.e., where the spacetime manifold is of the form $\mbb{R}\times T^{10}/\mbb{Z}_n$. The ``moduli space'' of M-theory on $T^{10}$ is described by the arithmetic quotient $E_{10}(\mbb{Z})\verb|\|E_{10}(\mbb{R})\verb|/|K(E_{10})$, and the orbifold projection reduces the global symmetry to a subgroup $\mc{G}\subset E_{10}$. A surprising feature of the analysis of \cite{Carlevaro} is that for some orbifolds the algebra $\mf{g}=\text{Lie}\ \mc{G}$ is a Borcherds subalgebra of $E_{10}$. We expect that a similar phenomenon would occur in our analysis by allowing for internal manifolds of the form $\mc{M}\verb|/|\mbb{Z}_n$. For example, projecting out gravity walls would expose subdominant walls associated with imaginary roots of $E_{10}$, which could give rise to billiards controlled by Borcherds algebras. Since the Weyl groups of Borcherds algebras are defined only with respect to the real simple roots, and the walls corresponding to non-real simple roots are lightlike, one might expect that chaos will always be removed for these cases.\footnote{We thank Axel Kleinschmidt for discussions on this issue.} We leave an investigation of this for future work.


\section*{Acknowledgements}

We thank Sophie de Buyl, Jarah Evslin, Gary Gibbons, Axel Kleinschmidt, Bengt E. W. Nilsson, Jakob Palmkvist, Malcolm Perry, Christoffer Petersson, Larus Thorlacius and Alexander Wijns for discussions at various stages in the preparation of this article. We are also very grateful to Jakob Palmkvist for carefully reading the manuscript and giving numerous useful suggestions. Finally, D.H.W. and D.P. would like to thank, respectively, Universit\'e Libre de Bruxelles and the Centre for Theoretical Cosmology at Cambridge University for generous hospitality while this work was carried out.

Work supported in part by IISN-Belgium
(convention 4.4511.06 (M.H.) and convention 4.4514.08 (M.H.
and D.P)), by the Belgian National Lottery, by the European
Commission FP6 RTN programme MRTN-CT-2004-005104 (M.H. and
D.P.), and by the Belgian Federal Science Policy Office through the
Interuniversity Attraction Pole P6/11.

\vspace{1cm}



\begin{appendix}

\section{The Complete Classification}

In this appendix we present the results obtained for all compactifications of the models associated with finite simple Lie groups. The $A_n$ series, representing pure gravity is not affected by the
compactifications we consider here and hence is not presented.

Note that the low-rank examples of the infinite families $B_n$ and $D_n$ must be treated separately.  This is because for compact orientable $N$-dimensional compactification manifolds we have
$b_j = b_{N-j}$
and so not all of the Betti numbers are independent. Therefore, we cannot set independently to zero the  Betti numbers relevant to the $p$-forms in the problem when the dimension of space is ``too small". Accordingly, for these infinite families we present the ``large-$n$" and ``small-$n$" cases in separate tables.

For each Einstein-$p$-form system and compactification, we give the Coxeter graph of the formal Coxeter group associated with the billiard. Obtuse angles between the dominant walls are indicated by circles around the corresponding nodes in the Coxeter graph. When the billiard table is a Coxeter simplex we give the Dynkin diagram.  Our wall conventions follow those of \cite{Sophie}. The symmetry walls are numbered such that the one corresponding to the overextended node has the highest power of the scale factors, i.e.,
\beq
\al_{-1}(\be)=\be^{n}-\be^{n-1}, \qquad \al_0(\be)=\be^{n-1}-\be^{n-2}, \dots \text{etc}.
\eeq
The non-symmetry dominant walls are given separately for each case in the forthcoming tables.

The computations in this section were carried out with the assistance of a custom computer program, which is freely available for download at: \\ http://www.damtp.cam.ac.uk/user/dhw22/code/.

\begin{table}
\begin{center}
\begin{tabular}{|m{12mm}|m{12mm}|m{50mm}|m{45mm}|m{14mm}|}
\hline
& & & &\\
$\{b_i(\mc{M})$ $=0\}$ & Coxeter group & Comments  &  Coxeter Graph of formal Coxeter group / Dynkin Diagram  & Chaotic? \\
& & & &\\
    \hline
     \hline
$\{\hs\},$ $\{b_1\},$ $\{b_2\},$ $\{b_1,b_2\}$ &    $E_6^{++}$ &\tiny{The uncompactified case. The oxidation endpoint is $D=8$ and comprises a dilaton $\phi$, an axion $\chi$ and a $3$-form $C$. This theory can be obtained as a truncation of eleven-dimensional supergravity \cite{PopeJulia}. The non-symmetry dominant walls are the electric walls of $C$ and $\chi$, $\al_{4}(\be)=\be^1+\be^2+\be^3-\phi/\sqrt{2}$ and $\al_5(\be)=\sqrt{2}\phi$, respectively \cite{Sophie}.}  &\includegraphics[width=35mm]{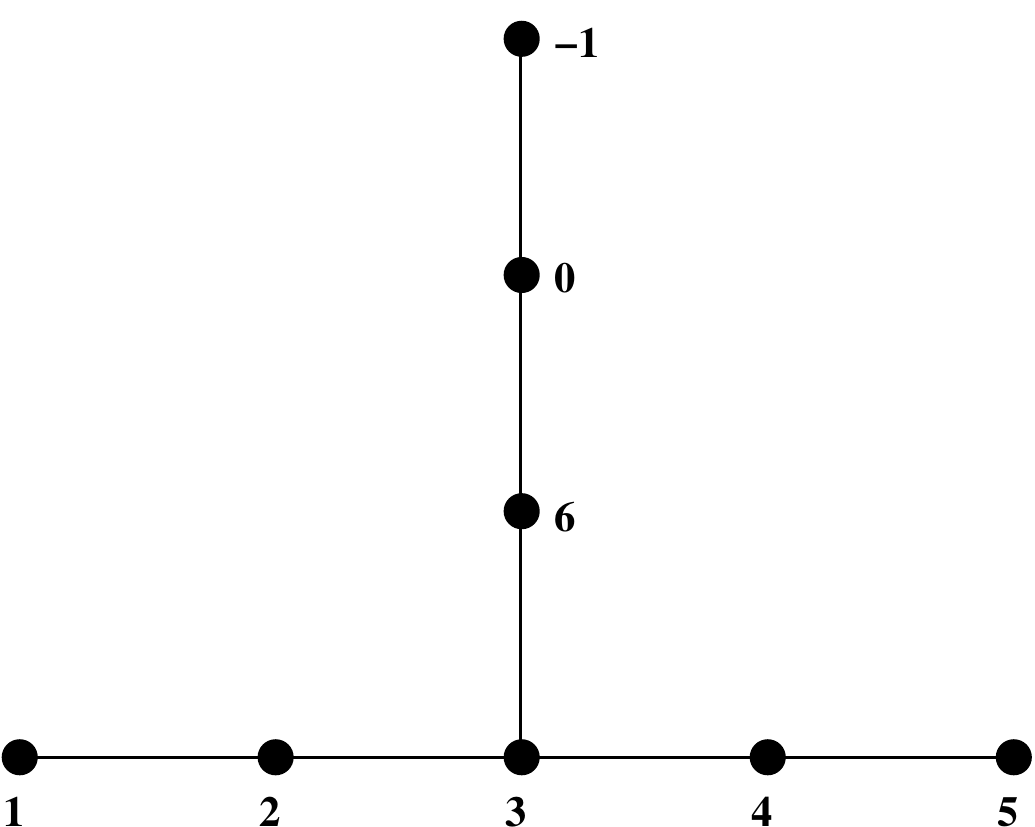}  & Yes \\
    \hline
$\{b_{3}\}$, $\{b_2, b_3\}$,    & $X_{9}$ & \tiny{In this compactification only the electric and magnetic walls of the axion $\chi$ are retained. Because the scalar products between $0$-form walls and gravity walls are always zero (see Section \ref{section:rules}), the axion walls and the gravity wall can peacefully coexist in this case while still giving rise to a Coxeter polyhedron, which is not, however, a simplex.  Its Gram matrix is degenerate. The Coxeter group $X_9$ is therefore a non-simplex Coxeter group in 7-dimensional hyperbolic space (9 faces). The matrix $\bar{A}$ is a valid Cartan matrix. However, this Cartan matrix is degenerate, and hence the associated Kac-Moody algebra $\tilde{\mf{g}}(\bar{A})$ is not simple \cite{Kac}. In fact, it contains a one-dimensional ideal $\mf{i}=\mbb{R}k$, where $k\in \mf{h}\subset E_6^{++}$ is the linear combination of Cartan generators of $E_6^{++}$ with coefficients given by the components of the null vector $v$ which spans the kernel of $\bar{A}$. See \cite{GeometricConf} for a detailed discussion of this phenomenon.   }&\includegraphics[width=47mm]{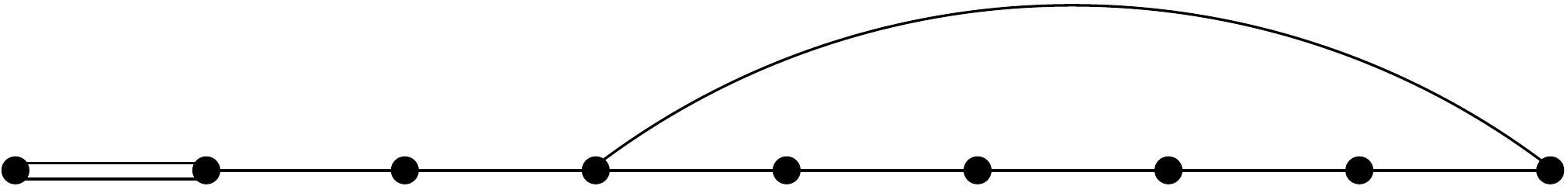} & Yes \\
        \hline
$\{b_1, b_3\}$, $\{b_1, b_2,$ $b_3\}$    & $A_5^{++}\times A_1$ &\tiny{Here, only the electric wall of the axion is retained, which has zero scalar product with the gravity and symmetry walls. Hence, the Coxeter group corresponds to the direct product of the ``pure gravity group'' $A_5^{++}$ and the group $A_1$ generated by reflections in the electric wall. The volume of the billiard region is infinite. }  & \includegraphics[width=45mm]{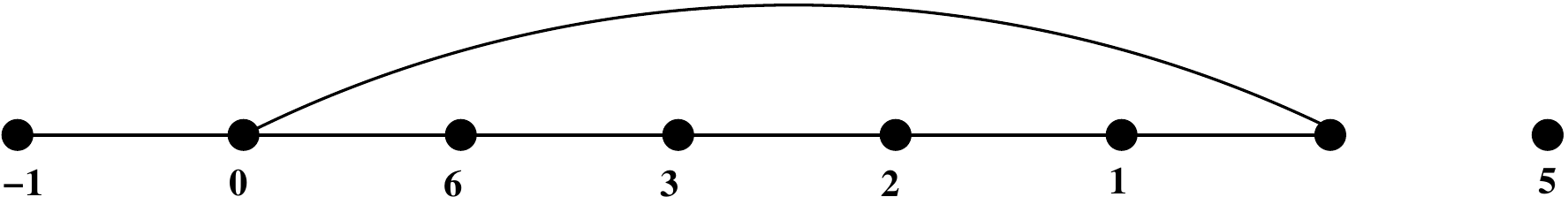} & No\\
        \hline
 \end{tabular}
         \caption{The $E_6$-sequence. The leftmost column indicates which Betti numbers vanish.}
\label{table:E6++}
\end{center}
\end{table}

\begin{table}
\begin{center}
\begin{tabular}{|m{12mm}|m{12mm}|m{50mm}|m{45mm}|m{14mm}|}
\hline
& & & & \\
$\{b_i(\mc{M})$ & Coxeter Group  &Comments  & Coxeter Graph of formal Coxeter group / Dynkin Diagram  & Chaotic? \\
$=0\}$ & & & &\\
& & & & \\
    \hline
     \hline
$\{\hs\}$, $\{b_2\}$, $\{b_1,b_2\}$ &    $E_7^{++}$ & \tiny{The uncompactified case. This theory oxidises to $D=9$ where it corresponds to a consistent non-supersymmetric truncation of maximal $D=9$ supergravity. The field content includes a dilaton $\phi$, and Maxwell field $A$ and a $3$-form $C$ \cite{PopeJulia}. The non-symmetry dominant walls are the electric walls of $A$ and $C$, $\al_{6}(\be)=\be^1-\f{2\sqrt{2}}{\sqrt{7}}\phi$ and $\al_7(\be)=\be^1+\be^2+\be^3+\f{\sqrt{2}}{\sqrt{7}}\phi$, respectively \cite{Sophie}. } &\includegraphics[width=45mm]{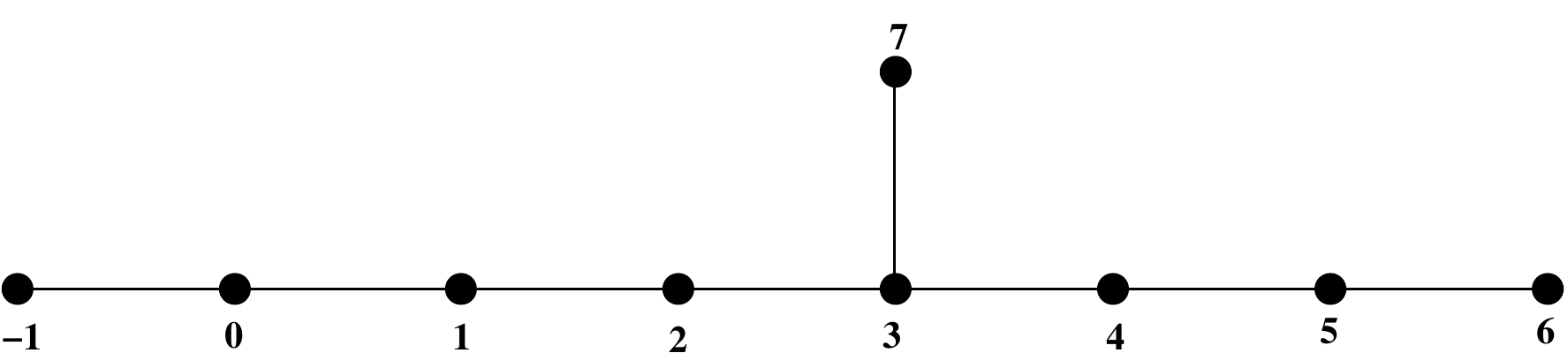}  & Yes \\
    \hline
 $\{b_{1}\}$    & $E_7^{++}$  & \tiny{Compactification on a manifold $\mc{M}$ with $b_1=0$ projects out the dominant wall $\al_6$ and promotes the magnetic wall form of the 3-form, $\tilde{\al}_6=m_{1234}^{[C]}$, to become dominant. The resulting set of wall forms does not define an acute-angled polyhedron. However, we may still associate a Coxeter graph to this wall system, which defines its formal Coxeter group $\mf{C}$. The group $\mf{C}$ has a normal subgroup $\mf{I}$, and taking the quotient by this subgroup yields the billiard Coxeter group $\mf{B}$, which turns out to be the original Coxeter group, $E_7^{++}=\mf{C}/\mf{I}$. Therefore the resulting dynamics is chaotic, even though the billiard region does not correspond to the fundamental Weyl chamber of $E_7^{++}$. See Section \ref{section:E7} for a detailed discussion of this compactification. }  &\includegraphics[width=45mm]{E7++b1.pdf} & Yes\\
      \hline
  $\{b_3\}$ & $E_7^{++}$ & \tiny{In this compactification the wall corresponding to the simple root $\al_7$ is projected out. There are two new dominant wall formS  One of them is again the magnetic wall form of the 3-form, $\bar{\al}_7= m_{1234}^{[C]}$. The other is the wall form $\tilde{\al}_7= m_{123456}^{[A]}$.  The billiard region is a finite union of images of the fundamental region of $E_7^{++}$ and hence the dynamics is chaotic. The formal Coxeter group $\mf{R}$ associated with the (non acute-angled) polyhedron defined by all dominant walls except $\tilde{\al}_7$ has a normal subgroup $\mf{J}$, and the quotient $\mf{R}/\mf{J}$ is again equal to $E_7^{++}$. See Section \ref{section:E7} for a detailed discussion of this compactification. }  & \includegraphics[width=45mm]{E7++b3.pdf} & Yes \\
  \hline
  $\{b_1, b_3\}$ & $D_6^{+++}$ &\tiny{This compactification is slightly subtle. Both the magnetic walls $m_{1234}^{[C]}$ and $m_{123456}^{[A]}$ are now dominant, together with the gravity wall $G_{178}$. Thus the polyhedron is not a simplex and furthermore, the gravity walls intersects the magnetic walls at obtuse angles. The Coxeter graph displayed corresponds to the wall system obtained by excluding the gravity wall. The Coxeter group $D_6^{+++}$ associated with this graph is not hyperbolic (although it is of course Lorentzian). However, we have explicitly checked that the new billiard region does not contain any spacelike rays, and hence the dynamcis is nevertheless chaotic. The reason is that the gravity wall comes into play and ``cuts'' the fundamental region of $Y_9$ in such a way that it shields the spacelike direction of escape and thus renders the total volume finite.  It is noteworthy that a triple-extended Kac-Moody algebra plays a role in the billiard structure.}   &\includegraphics[width=45mm]{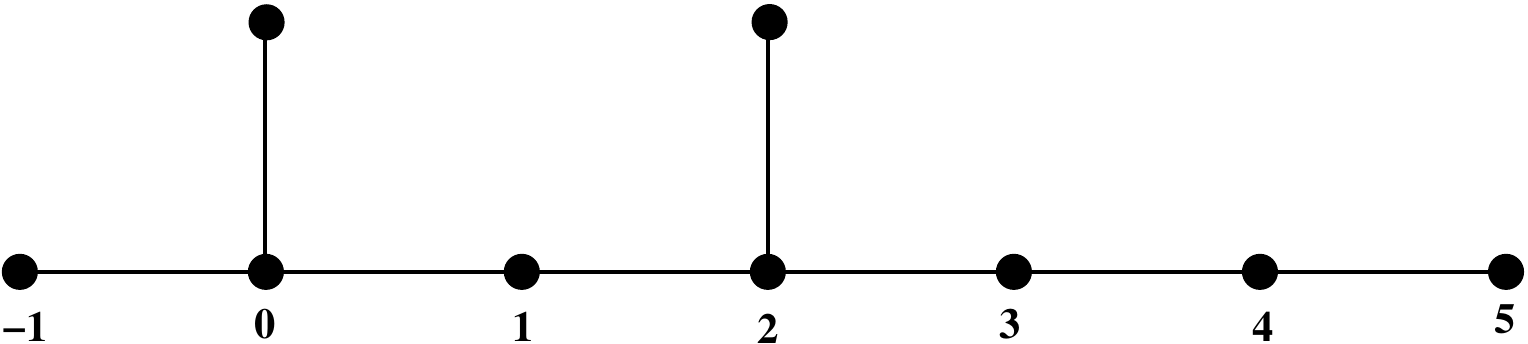} & Yes \\
  \hline
 \end{tabular}
         \caption{The $E_7$-sequence.}
\label{table:E7++1}
\end{center}
\end{table}

\begin{table}
\begin{center}
\begin{tabular}{|m{12mm}|m{12mm}|m{50mm}|m{45mm}|m{14mm}|}
\hline
& & & & \\
$\{b_i(\mc{M})$ & Coxeter Group  &Comments  & Coxeter Graph of formal Coxeter group / Dynkin Diagram   & Chaotic? \\
$=0\}$ & & & &\\
  \hline
  $\{b_2, b_3\}$ & $E_7^{++}$ &\tiny{The compactification with $b_2=b_3=0$ gives rise to the same Coxeter group structure as the $b_3=0$ case. The additional constraint $b_2=0$ removes the magnetic wall $m_{123456}^{[A]}$, while preserving the magnetic wall $m_{1234}^{[C]}$ and promoting the gravity wall $G_{178}$ to a dominant wall. Thus, excluding the gravity wall, we find the formal Coxeter group $\mf{R}$. The quotient group $\mf{R} / \mf{J}$ is again $E_7^{++}$, and the resulting dynamics is chaotic.} & \includegraphics[width=45mm]{E7++b3.pdf} & Yes \\
        \hline
  $\{b_1, b_2,$ $b_3, b_4\}$ & $A_6^{++} $ & \tiny{In this compactification, all $p$-form walls are projected out. In the absence of the dilaton (pure gravity), we would obtain a chaotic system. However, because the oxidation endpoint of the $E_7$-sequence includes a dilaton, which is never projected out, the resulting dominant wall system is not sufficient to provide a finite volume billiard. This can be seen from the fact that there are eight dominant walls, corresponding to the eight simple roots of $A_6^{++}$, while the total space in which the dynamics takes place is nine-dimensional. Hence, even though $A_6^{++}$ is hyperbolic, the billiard domain is of infinite volume since there is always an extra ``dimension'' in which the particle may escape to infinity. The compactification $b_1=b_2=b_3$ gives rise to the same dynamics since the magnetic wall which is preserved is not sufficient to provide a finite volume billiard.} & \includegraphics[width=45mm]{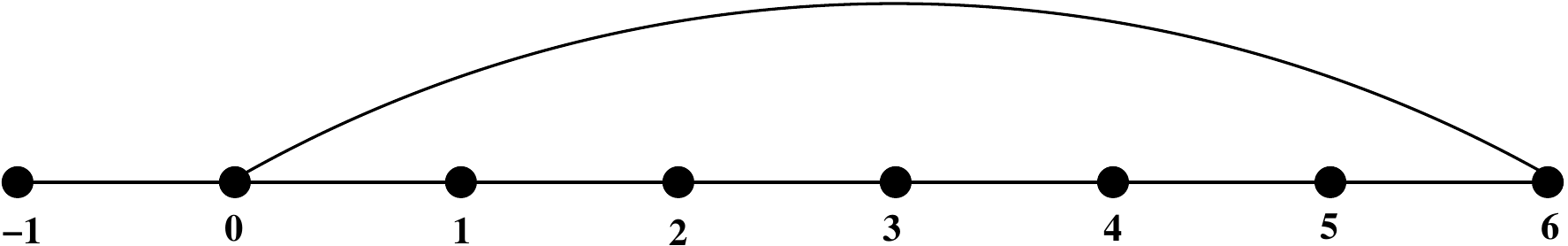} & No \\
  \hline
 \end{tabular}
         \caption{The $E_7$-sequence continued.}
\label{table:E7++2}
\end{center}
\end{table}

\begin{table}
\begin{center}
\begin{tabular}{|m{12mm}|m{12mm}|m{50mm}|m{45mm}|m{14mm}|}
\hline
& & & &\\
$\{b_i(\mc{M})$ $=0\}$ & Coxeter group & Comments & Coxeter Graph of formal Coxeter group / Dynkin Diagram  & Chaotic? \\
& & & & \\
\hline
\hline
$\{\hs \}$ &    $E_8^{++}$ & \tiny{The uncompactified case. The oxidation endpoint is eleven-dimensional supergravity, containing a $3$-form potential $C$ \cite{PopeJulia}. The only non-symmetry dominant wall is the electric wall of $C$, $\al_{8}(\be)=\be^1+\be^2+\be^3$ \cite{Sophie}.}& \includegraphics[width=47mm]{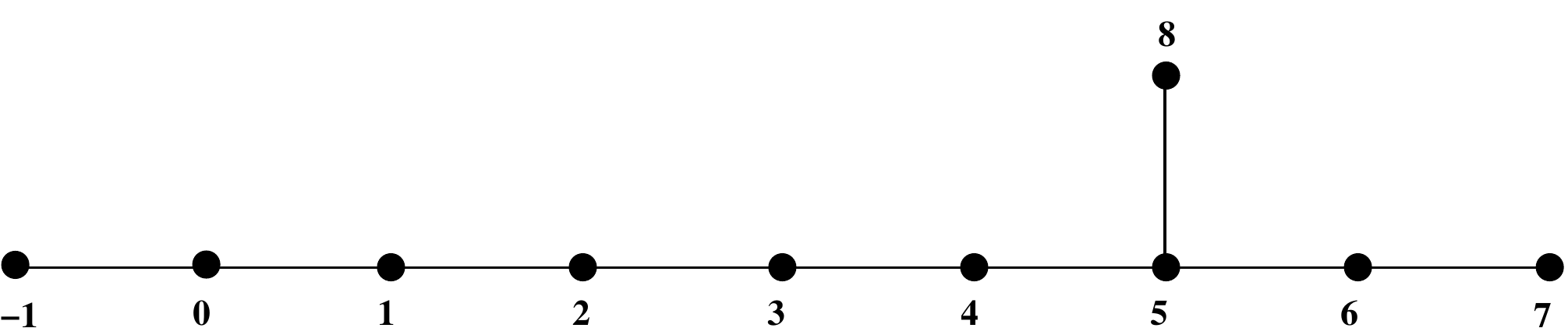}  & Yes \\
    \hline
 $\{b_3\}$    & $E_7^{+++} $  & \tiny{This compactification renders the magnetic wall of the 3-form dominant, while the electric wall is projected out. However, also the gravity wall becomes dominant, implying that the billiard table is not a simplex, and has obtuse angles. Excluding the gravity wall, the reflections in the new dominant walls give rise to the (non-hyperbolic) Lorentzian Coxeter group $E_7^{+++}$. The fundamental Weyl chamber of $E_7^{+++}$ is of infinite volume, which would imply non-chaotic dynamics. However, here the gravity wall again comes in and shields the direction of escape, rendering the total volume finite and preserves the chaotic dynamcis. It is noteworthy that a triple-extended Kac-Moody algebra plays again a role in the billiard structure. }&\includegraphics[width=45mm]{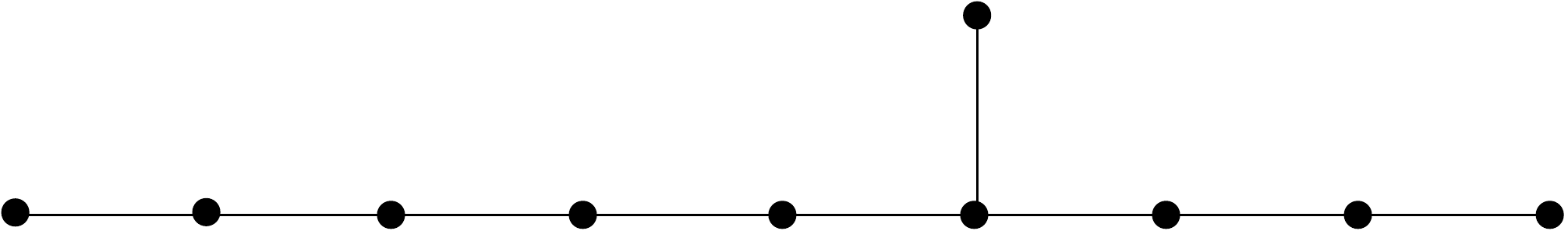} & Yes \\
      \hline
 $\{b_3,b_4\}$  & $A_8^{++}$ &\tiny{This corresponds to the standard scenario when all of the $p$-form walls are projected out. The billiard is the same as for pure gravity in eleven dimensions, which is non-chaotic since the Kac-Moody algebra $A_8^{++}$ is not hyperbolic.}   &  \includegraphics[width=47mm]{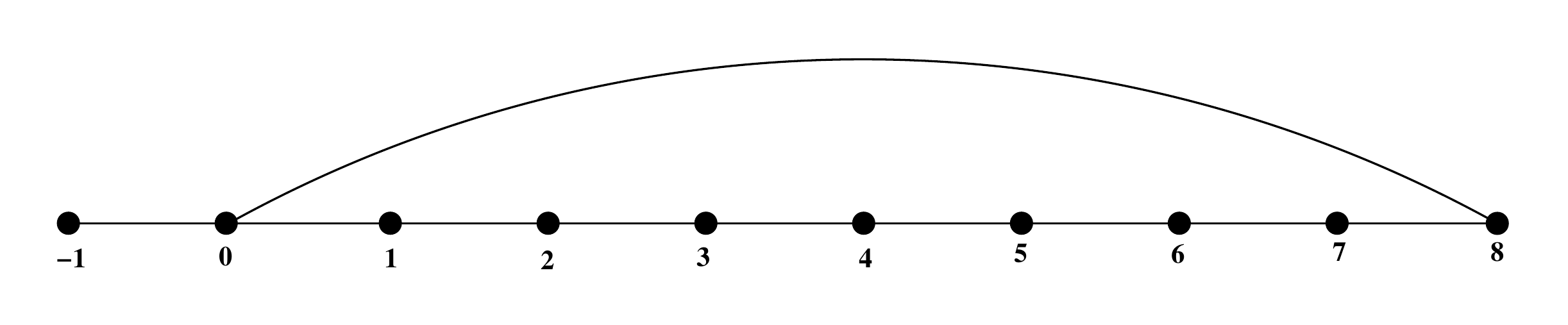} & No \\
   \hline
 \end{tabular}
         \caption{The $E_8$-sequence.}
\label{table:E8}
\end{center}
\end{table}

\begin{table}
\begin{center}
\begin{tabular}{|m{12mm}|m{12mm}|m{50mm}|m{45mm}|m{14mm}|}
\hline
& & & & \\
$\{b_i(\mc{M})$ $=0\}$ & Coxeter Group & Comments   & Coxeter Graph of formal Coxeter group / Dynkin Diagram & Chaotic? \\
& & & &\\
    \hline
     \hline
$\{\ \}, \{b_2\},$ $\{b_3\}$, $\{b_2, b_3\}$ &    $F_4^{++}$ &\tiny{The uncompactified case. The oxidation endpoint is $D=6$ and the field content is given by a dilaton $\phi$, and axion $\chi$, two Maxwell fields $A^{\pm}$, a $2$-form $B$ and a $3$-form potential $C$ with self-dual field strength $G$ \cite{PopeJulia}. The non-symmetry dominant walls are the electric walls of $A^{-}$ and $\chi$, $\al_3(\be)=\be^1-\phi/2$ and $\al_4(\be)=\phi$, respectively \cite{Sophie}. } & \includegraphics[width=45mm]{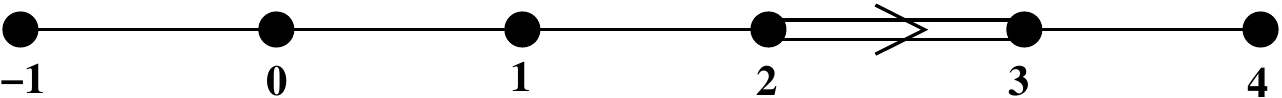}  & Yes \\
    \hline
 $\{b_{1}\},$ $\{b_1, b_3\}$    & $B_4^{++}$  & \tiny{This compactification renders the one-form $A^{-}$ massive and hence projects out the associated electric wall $\om_{3}(\be)=\be^{1}-\phi/2$. The new dominant wall is the electric wall of the 2-form $B$, which reads $\tilde{\om}_3(\be)=\be^1+\be^2-\phi$. The new dominant wall may be identified with the simple roots of the hyperbolic Kac-Moody algebra $B_4^{++}$, and the reflections in the faces of these walls generate the Weyl group $\mf{W}[B_4^{++}]$. The billiard domain $\mc{B}$ coincides with the fundamental Weyl chamber $\mc{F}$ of $\mf{W}[B_4^{++}]$ and since this is of finite volume, the dynamics is chaotic. This case is discussed in more detail in Section \ref{section:F4}.} &\includegraphics[width=45mm]{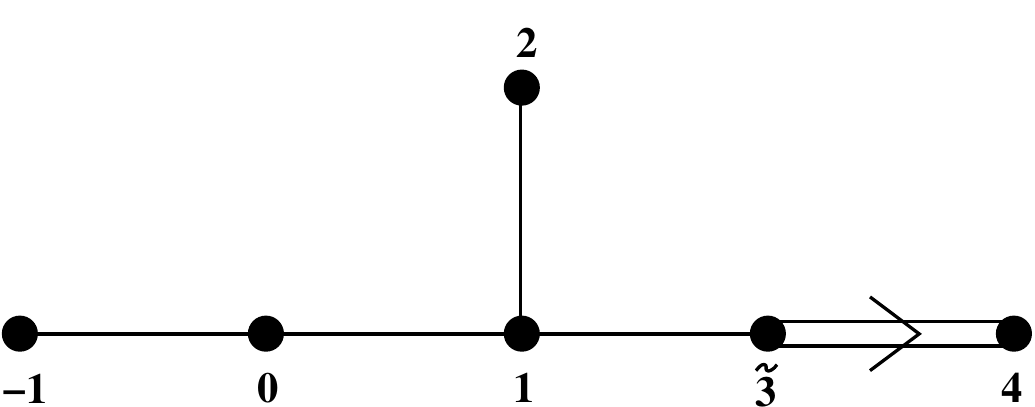} & Yes \\
      \hline
  $\{b_1, b_2\}$, $\{b_1, b_2,$  $b_3\}$ & $A_3^{++}\times A_1$ & \tiny{In this compactification, all $p$-form walls are projected out, but the axion wall $\om_{4}=\phi$ is always present. Therefore the resulting dominant wall system is the standard one $A_3^{++}$ of pure gravity in 6 dimensions, augmented with the extra axion wall. Even though $A_3^{++}$ is hyperbolic, there exists a spacelike coweight due to the finite $A_1$-factor, and therefore the dynamics is non-chaotic. This case is discussed in more detail in Section \ref{section:F4}.}  &  \includegraphics[width=45mm]{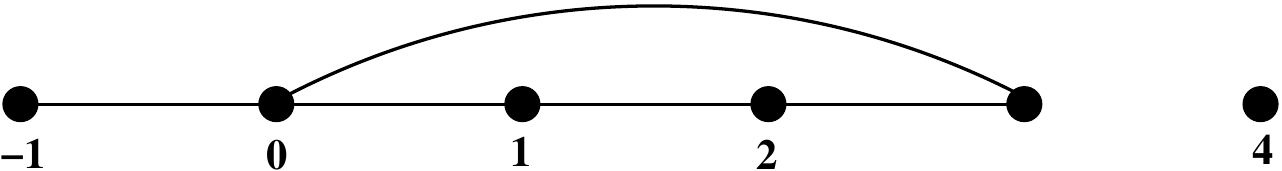} & No \\
   \hline
 \end{tabular}
         \caption{The $F_4$-sequence.}
\label{table:F4++}
\end{center}
\end{table}

\begin{table}
\begin{center}
\begin{tabular}{|m{12mm}|m{12mm}|m{50mm}|m{45mm}|m{14mm}|}
\hline
& & & & \\
$\{b_i(\mc{M})$ $=0\}$ & Coxeter Group   &  Comments  & Coxeter Graph of formal Coxeter group / Dynkin Diagram  & Chaotic? \\
& & & & \\
    \hline
     \hline
$\{ \}, \{b_2\},$ $\{b_3\}$ &    $G_2^{++}$ &\tiny{The uncompactified case. The oxidation endpoint is $5$-dimensional Maxwell-Einstein gravity \cite{PopeJulia}. The connection between this supergravity theory and $G_2^{++}$ has also been extensively investigated in \cite{Mizoguchi2}. The only non-symmetry dominant wall is the electric wall of the Maxwell field $A$, $\al_2(\be) = \beta^1$ \cite{Sophie}.  } & \includegraphics[width=40mm]{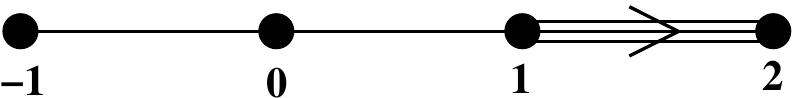}  & Yes \\
    \hline
 $\{b_{1}\}$   & $A_2^{++}$ & \tiny{This compactification simply renders the one-form $A$ massive and thus projects out the electric wall $\om_2(\be)=\be^{1}$. The new dominant wall is the gravity wall $\tilde{\om}_2(\be)=G_{134}(\be)=2\be^{1}+\be^2$, and not the magnetic wall $m^{[A]}_{123}(\be)=\be^{1}+\be^{2}+\be^{3}$, since we have $m^{[A]}_{123}(\be)=G_{134}(\be)+\om_0(\be)+\om_1(\be)$ (see \cite{Sophie} for our wall conventions). The billiard dynamics is controlled by the Weyl group of the hyperbolic Kac-Moody algebra $A_2^{++}$, and is therefore chaotic.}  & \includegraphics[width=40mm]{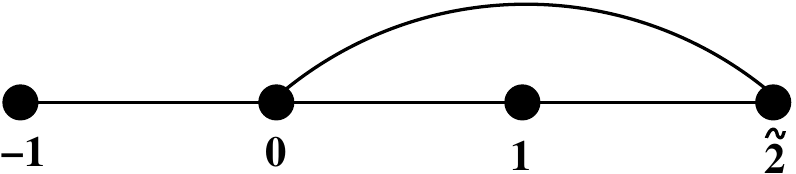} & Yes\\
        \hline
 \end{tabular}
         \caption{The $G_2$-sequence.}
\label{table:G2++}
\end{center}
\end{table}

\begin{table}
\begin{center}
\begin{tabular}{|m{12mm}|m{12mm}|m{45mm}|m{50mm}|m{14mm}|}
\hline
& & & & \\
$\{b_i(\mc{M})$ $=0\}$ &Coxeter Group   &Comments  & Coxeter Graph of formal Coxeter group / Dynkin Diagram & Chaotic? \\
& & & &\\
    \hline
     \hline
$\{\hs\},\{b_2\}$ &    $B_n^{++}$ &\tiny{The uncompactified case. The oxidation endpoint is in $D=n+2$ dimensions and includes a dilaton $\phi$, a $2$-form $B$ and a Maxwell field $A$ \cite{PopeJulia}. The non-symmetry dominant walls are the electric wall of the Maxwell field $A$ and the magnetic wall of the $2$-form $B$, $\al_n(\be)=e_1^{[A]}(\be)=\be^{1}+\f{a}{2\sqrt{2}}\phi$ and $\al_1(\be)=m_{1\cdots (n-2)}(\be)=\be^{1}+\cdots +\be^{n-2}-\f{a}{\sqrt{2}}\phi$ ($a^{2}=8/n$) \cite{Sophie}.} & \includegraphics[width=53mm]{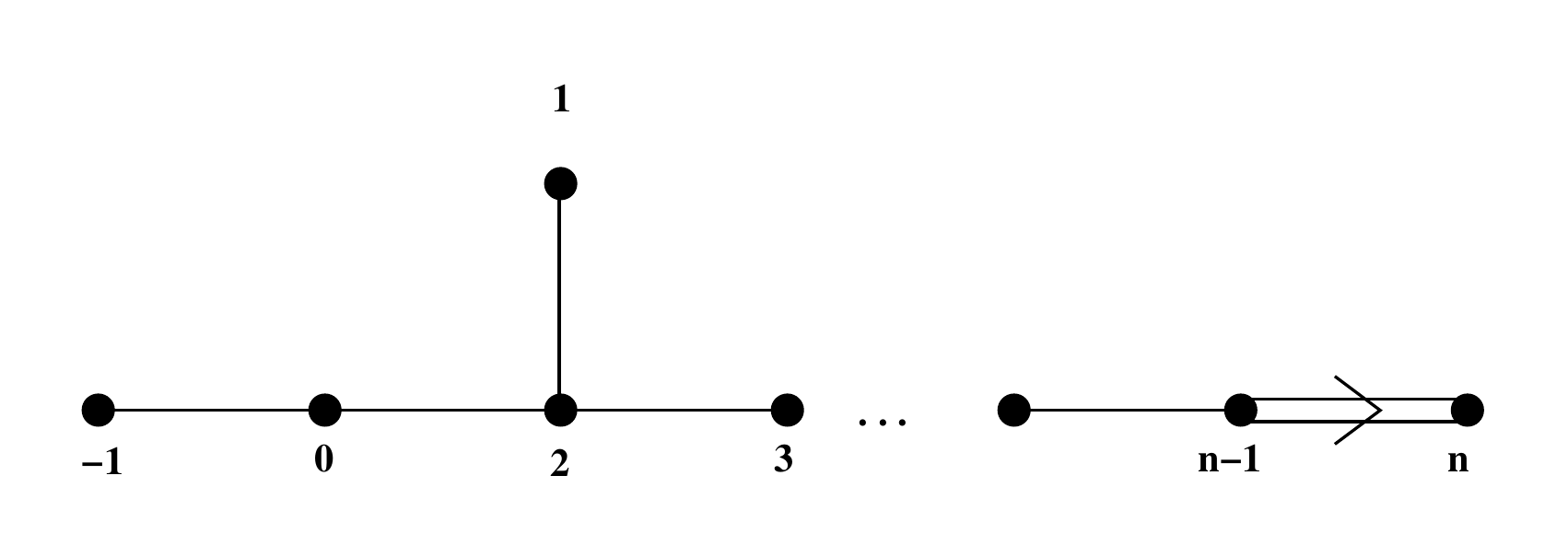}  & Yes for $n\le 8$, no for $n>8$.\\
        \hline
$\{b_1\}$ &    $D_n^{++}$ &\tiny{This compactification projects our the electric wall $e^{[A]}_1(\be)$ corresponding to node $n$ in the original diagram. The new dominant wall is the electric wall of the $2$-form $e^{[B]}_{12}(\be)$ which attaches by a single link to node $n-2$. The resulting Kac-Moody algebra is $D_n^{++}$ which is hyperbolic for $1\leq n\leq 8$.} & \includegraphics[width=50mm]{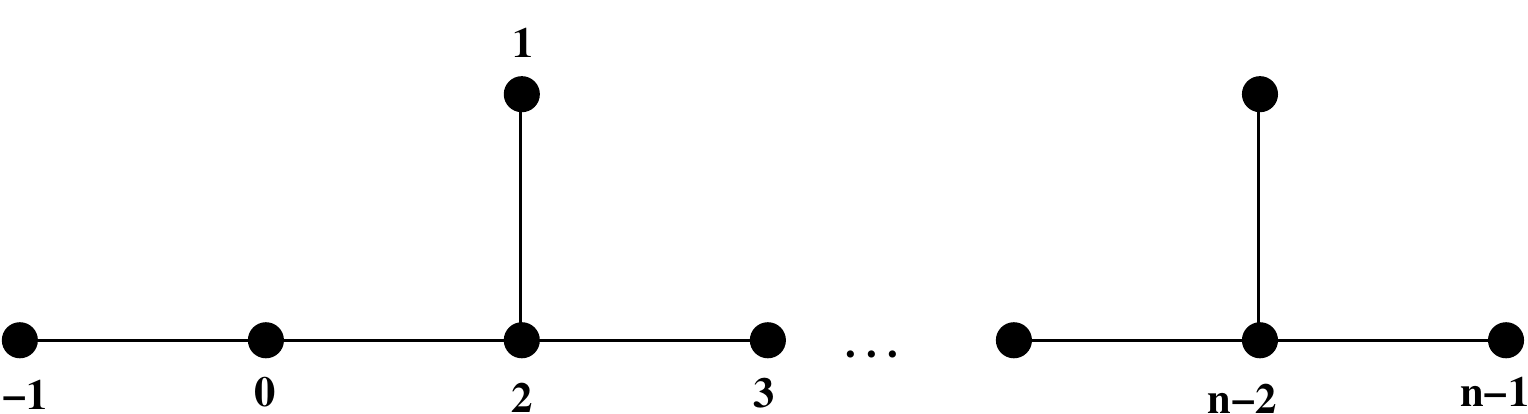}  & Yes for $n\le 8$, no for $n>8$.\\
        \hline
$\{b_3\}$ &    $R_{n+2}$ &\tiny{Here we project out the magnetic wall $m^{[B]}_{12\cdots (n-2)}(\be)$ of the $2$-form, corresponding to node number $1$ in the original diagram. The new dominant wall is the magnetic wall $m^{[A]}_{12\cdots (n-1)}(\be)$ which attaches by a double link to node $0$. The matrix of scalar products between the dominant walls is a valid Cartan matrix corresponding to the rank $n+2$ Lorentzian Kac-Moody algebras $R_{n+2}$. To our knowledge, this class of algebras is not part of any previous classification. It is an extension of the twisted affine algebra $D_{n+1}^{(2)}$ and fulfills the property of \cite{GOW}. Since the algebras are not hyperbolic, the dynamics is always non-chaotic.} &\includegraphics[width=53mm]{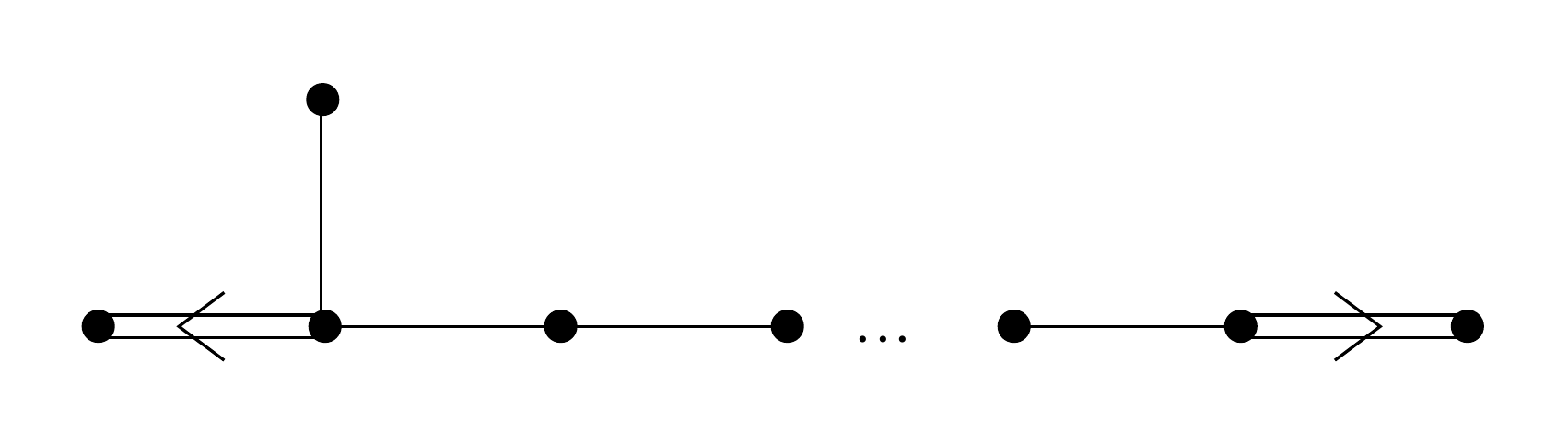}  & No\\
        \hline
$\{b_1,b_2\}$ &    $Z_{n+2} $ &\tiny{The electric wall $e^{[A]}_{1}(\be)$ of the Maxwell field and the electric wall $e^{[B]}_{12}(\be)$ of the $2$-form are both projected out. As a result the magnetic wall $m^{[B]}_{12\cdots (n-2)}(\be)$ and the gravity wall $G_{1n(n+1)}(\be)$ become dominant. Because of the facts explained in Section \ref{section:rules} the resulting billiard table is not acute-angled (but is a simplex). We have displayed the Coxeter graph of the formal Coxeter group associated to the billiard table. It is a (non-hyperbolic) Lorentzian Coxeter group, which we call $Z_{n+2}$. The billiard group is a quotient of the formal Coxeter group by the normal subgroup generated by the extra relations, which we have not worked out explicitly. We have checked that the volume of the billiard region is infinite so that the dynamics is not chaotic.}  &\includegraphics[width=50mm]{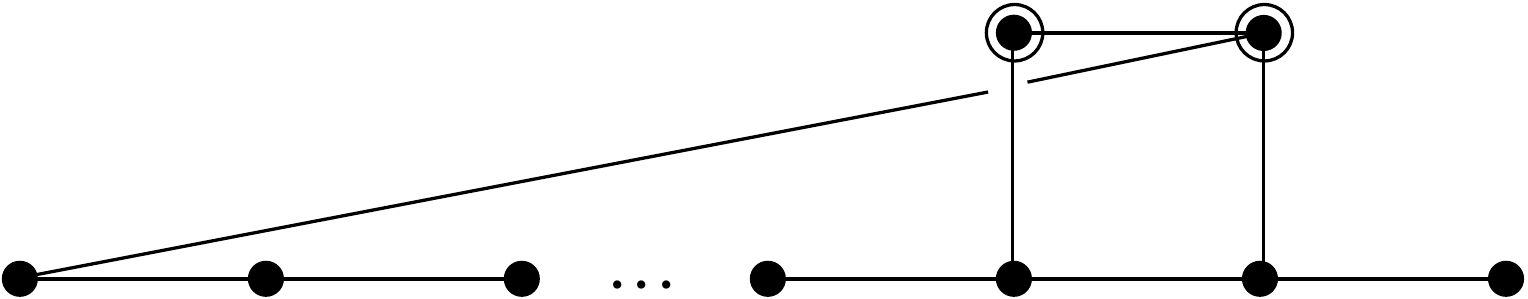}  & No\\
        \hline
 \end{tabular}
         \caption{The $B_n$-sequence for $n>4$.}
\label{table:Bn1}
\end{center}
\end{table}

\begin{table}
\begin{center}
\begin{tabular}{|m{12mm}|m{12mm}|m{45mm}|m{50mm}|m{14mm}|}
\hline
& & & & \\
$\{b_i(\mc{M})$ $=0\}$ &Coxeter Group   &Comments  & Coxeter Graph of formal Coxeter group / Dynkin Diagram  & Chaotic? \\
& & & &\\
    \hline
     \hline
$\{b_1,b_3\}$ &    ${S}_{n+2} $ & \tiny{This compactification displays an example of a ``coexistence'' of an electric, a magnetic and a gravity wall in the set of dominant walls. These correspond to $e^{[B]}_{12}(\be)$, $m^{[A]}_{12\cdots (n-1)}(\be)$ and $G_{1n(n+1)}(\be)$. The billiard table is therefore not a simplex. To analyze the situation we may consider the set of dominant walls with the gravity wall excluded. Then the matrix of scalar products between these walls gives rise to the Cartan matrix of a rank $n+2$ Lorentzian Kac-Moody algebra $S_{n+2}$, whose Dynkin diagram is displayed to the right. This is a non-standard extension of the affine algebra $B_n^{+}$. The fundamental domain for this algebra has infinite volume, and we have checked that including the gravity wall is not enough to render the billiard region finite. Thus, the dynamics is non-chaotic.}
&\includegraphics[width=53mm]{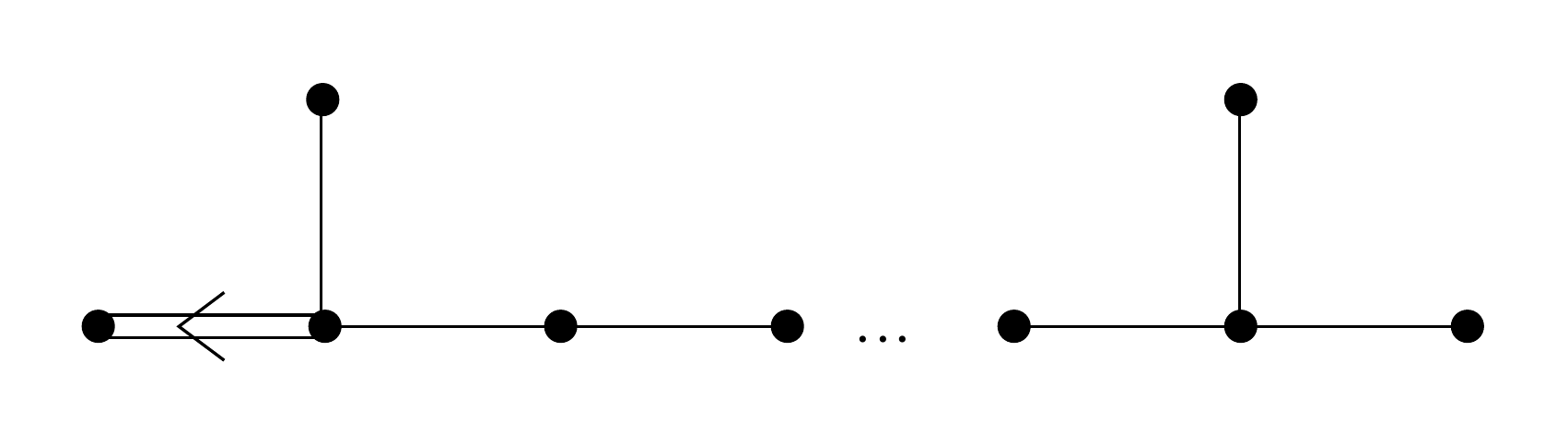}  & No\\
        \hline
$\{b_2,b_3\}$ &    ${Q}_{n+2} $ & \tiny{In this compactification the magnetic wall $\al_1(\be)=m^{[B]}_{1\cdots (n-2)}(\be)$ of the $2$-form $B$ is projected out and is replaced by the gravity wall $G_{1n(n+1)}(\be)$. The number of dominant walls is therefore $n+2$ and the billiard table is a simplex. However, due to the presence of the gravity wall it is not acute-angled. We display the Coxeter graph of the associated formal Coxeter group, denoted by $Q_{n+2}$. The region bounded by the new dominant walls contains spacelike coweights and hence the dynamics is non-chaotic.} &\includegraphics[width=50mm]{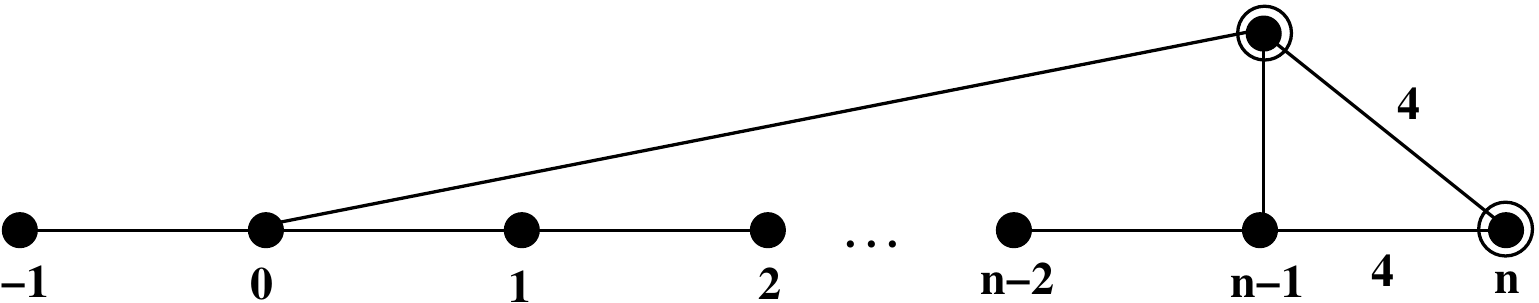}  & No\\
        \hline
$\{b_1,b_2,$ $b_3\}$ &    ${A}_{n-1}^{++}$  &\tiny{This compactification projects out all $p$-form walls, and so only the gravity and symmetry walls remain, giving rise to the rank $n-1$ Kac-Moody algebra $A_{n-1}^{++}$ which is hyperbolic for $2\leq n\leq 8$. Because of the presence of a dilaton, the total space $\mf{M}_{\be}$ is however $n+2$-dimensional and hence there is always a direction of escape, rendering the dynamics non-chaotic.} &\includegraphics[width=53mm]{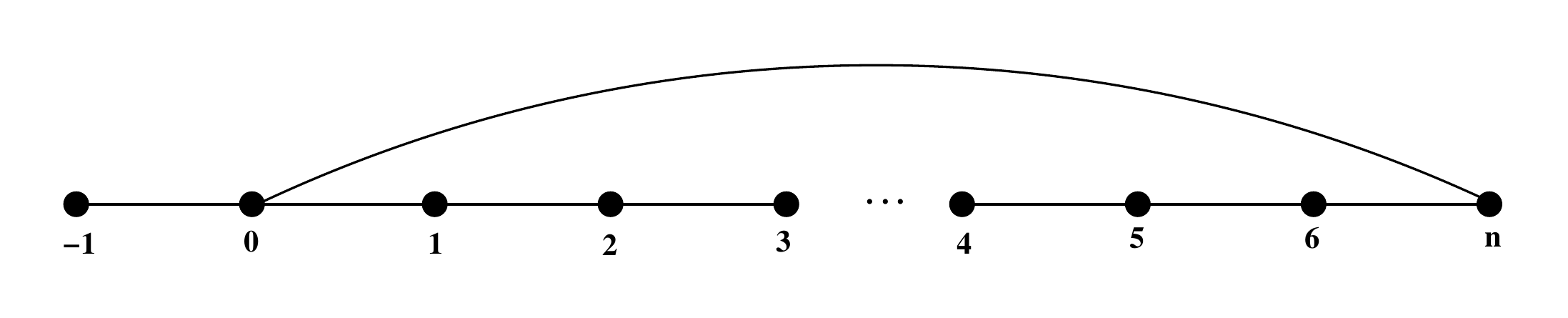} & No\\
        \hline
 \end{tabular}
         \caption{The $B_n$-sequence for $n>4$.}
\label{table:Bn2}
\end{center}
\end{table}

\begin{table}
\begin{center}
\begin{tabular}{|m{12mm}|m{12mm}|m{45mm}|m{50mm}|m{14mm}|}
\hline
& & & &\\
$\{b_i(\mc{M})$ $=0\}$ & Coxeter Group   &Comments & Coxeter Graph of formal Coxeter group / Dynkin Diagram   & Chaotic? \\
& & & & \\
    \hline
     \hline
$\{\}$ &    $C_n^{++}$ & \tiny{The uncompactified case. The oxidation endpoint corresponds is gravity in $D=4$ coupled to a collection of dilatonic scalars, axions and Maxwell fields \cite{PopeJulia}. See \cite{Sophie} for the definition of the dominant walls in this case.  } & \includegraphics[width=53mm]{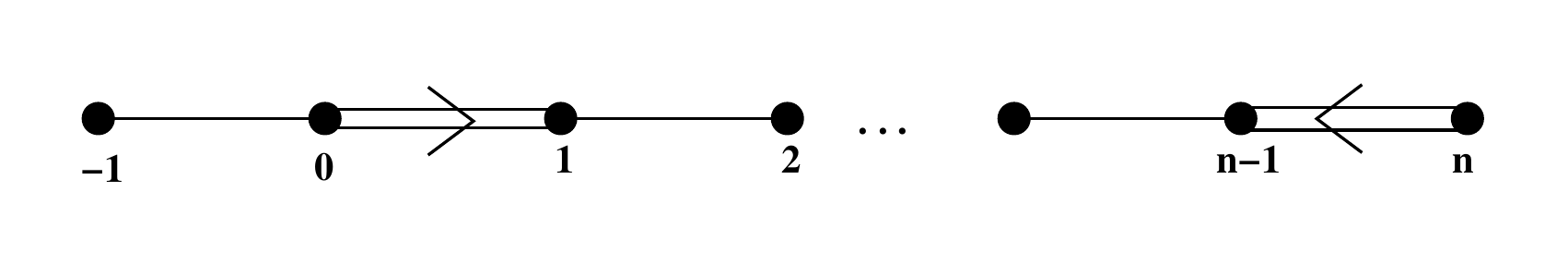}  & Yes for $n\leq 4$, no for $n>4$\\
        \hline
$\{b_1=$ $b_2\}$ &    $A_1^{++}$ $ \times\ C_{n-1}$ & \tiny{Because of Poincar\'e duality we have $b_1=b_2$ so the only relevant compactification is the one for which $b_1=b_2=0$. This renders all Maxwell fields massive, and the resulting theory is described by two disconnected diagrams: the pure gravity piece $A_1^{++}$, and a piece $C_{n-1}$ associated with the axions. Beacuse of the disconnected part corresponding to a finite Lie algebra we know that there exist spacelike coweights, and hence the dynamics is non-chaotic.} &\includegraphics[width=50mm]{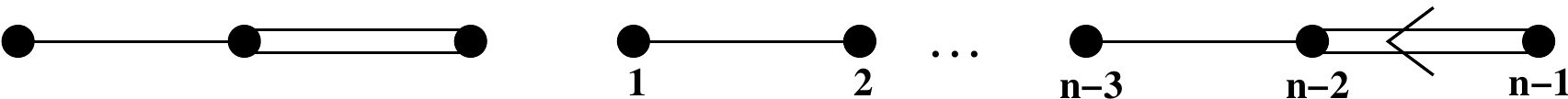}
 & No for $n\geq 2$ \\
        \hline
 \end{tabular}
         \caption{The $C_n$-sequence.}
\label{table:Cn}
\end{center}
\end{table}

\begin{table}
\begin{center}
\begin{tabular}{|m{12mm}|m{12mm}|m{45mm}|m{50mm}|m{14mm}|}
\hline
& & & &\\
$\{b_i(\mc{M})$ $=0\}$ & Coxeter Group & Comments  & Coxeter Graph of formal Coxeter group / Dynkin Diagram  & Chaotic? \\
& & & &\\
    \hline
     \hline
$\{\}, \{b_1\}$ &    $D_n^{++}$ &\tiny{The uncompactified case. The oxidation endpoint is $D=n+2$, with matter content given by a dilaton $\phi$ and a $2$-form $B$. The associated electric and magnetic walls, $e^{[B]}_{12}(\be)$ and $m^{[B]}_{12\cdots (n-2)}(\be)$, are both dominant before compactification \cite{PopeJulia}. The non-symmetry dominant walls are the electric and magnetic wall sof the $2$-form $B$, $\al_n(\be)=\be^{1}+\be^{2}+\f{a}{\sqrt{2}}\phi$ and $\al_1(\be)=\be^{1}+\cdots + \be^{n-2}-\f{a}{\sqrt{2}}\phi$, respectively ($a^2=8/n$) \cite{Sophie}.  } & \includegraphics[width=50mm]{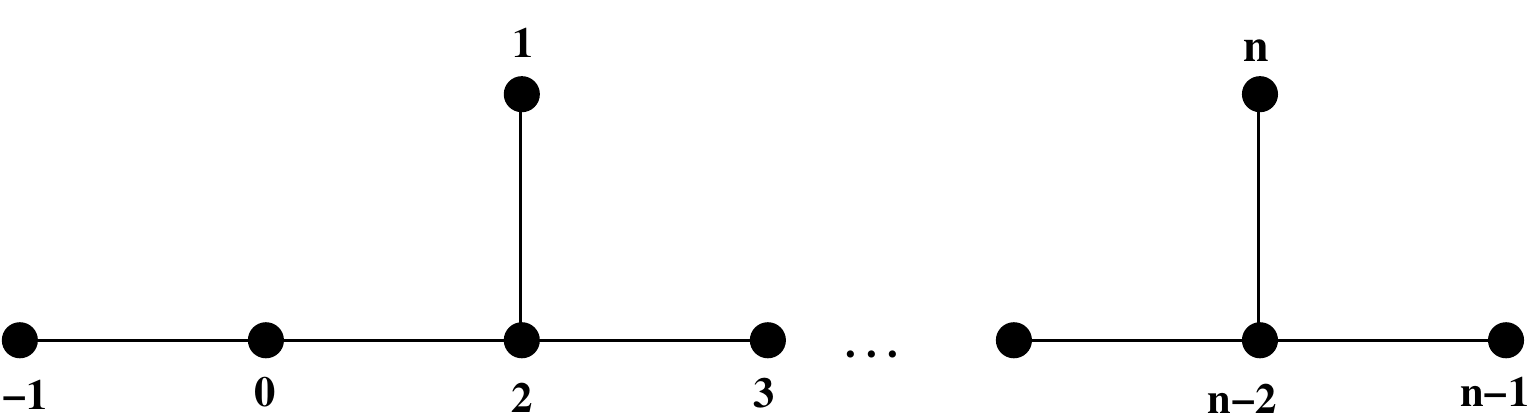}  & Yes for $n\le 8$, no for $n>8$.\\
        \hline
$\{b_2\},$ $\{b_1,b_2\}$ &    $Z_{n+2}$ & \tiny{Compactification on a manifold $\mc{M}$ with $b_2=0$ (or $b_1=b_2=0$) renders the electric field massive, and the electric wall is replaced by the gravity wall $G_{1n(n+1)}(\be)$. This new set of dominant walls do not define an acute-angled billiard table.  The formal Coxeter group associated with the reflections in these walls is again the Lorentzian Coxeter group $Z_{n+2}$, whose Coxeter graph is displayed to the right. This is the same Coxeter group which appears in the billiard of the $b_1=b_2=0$ compactification of the $B_n$-theory. The billiard region contains spacelike coweights and hence the dynamics is non-chaotic.} & \includegraphics[width=50mm]{Zn.pdf}  & No\\
        \hline
$\{b_3\},$ $\{b_1,b_3\}$ &    $W_{n+2}$ &\tiny{In this compactification the magnetic field becomes massive and hence the magnetic wall $m^{[B]}_{12\cdots (n-2)}(\be)$ is projected out, again rendering the gravity wall $G_{1n(n+1)}(\be)$ dominant. The new set of dominant walls does not define an acute-angled billiard table. The Coxeter graph derived from the dominant walls is displayed to the right and it corresponds to a (non-hyperbolic) Lorentzian Coxeter group. We have checked that the fundamental chamber contains spacelike coweights and hence the dynamics is non-chaotic.} & \includegraphics[width=45mm]{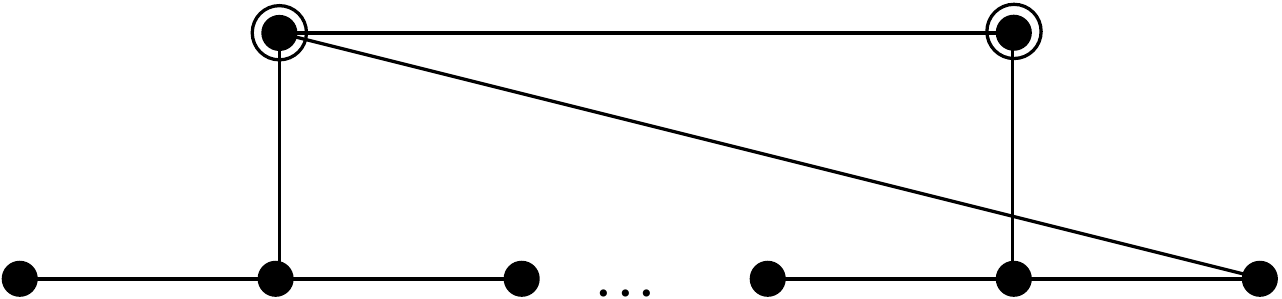}  & No\\
        \hline
$\{b_2,b_3\}$, $\{b_1,b_2,$ $b_3\}$ &    $A_{n-1}^{++}$ & \tiny{All $p$-form walls are projected out and the dominant wall system is the standard one $A_{n-1}^{++}$ for pure gravity. However, since there is a dilaton the total scale-factor space $\mf{M}_{\be}$ is $n+2$-dimensional, and there will always exist a direction of escape, rendering the theory non-chaotic. }& \includegraphics[width=50mm]{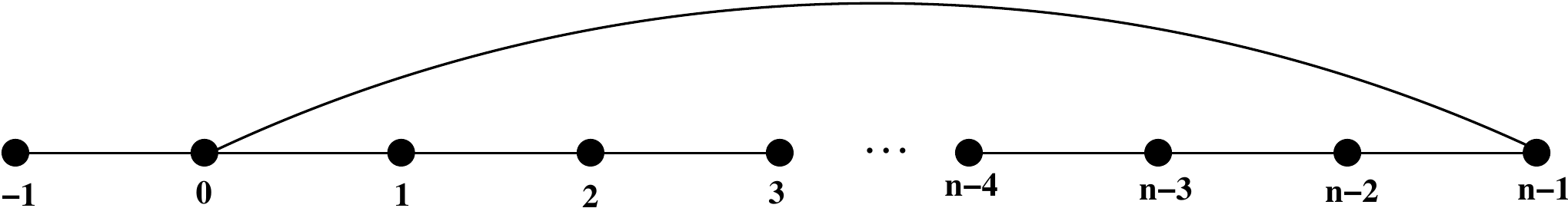}  & No\\
        \hline
 \end{tabular}
         \caption{The $D_n$-sequence for $n>4$.}
\label{table:Dn}
\end{center}
\end{table}

\begin{table}
\begin{center}
\begin{tabular}{ |m{25mm}|m{18mm}|m{18mm}|m{18mm}|}
\hline
$\{b_i(\mc{M})=0\}$ & $n=2$ & $n=3$ & $n=4$ \\
\hline\hline
$\{\},\{b_2\}$ & $B_2^{++}$ \copyright & $B_3^{++}$  \copyright & $B_4^{++}$ \copyright  \\
$\{b_1\}$ & $D_2^{++}$ \copyright & $D_3^{++}$ \copyright & $D_4^{++}$ \copyright\\
$\{b_1,b_2\}$ & $A_1^{++}\times A_1$ & $A_3^{++}$ \copyright & $A_3^{++}$ \\
$\{b_3\}$ & $K_4$ \copyright &$A_3^{++}$ \copyright  & $L_6$ \\
$\{b_1,b_3\}$ & $A_1^{++}$ &$A_3^{++}$ \copyright  & $A_3^{++}$ \\
$\{b_2,b_3\}$ & $A_1^{++}$          &$A_3^{++}$ \copyright  & $L_6$ \\
$\{b_1,b_2,b_3\}$ &     $A_1^{++}$       &$A_3^{++}$ \copyright  & $A_3^{++}$ \\
\hline
\end{tabular}
\caption{Compactifications of the $B_n^{++}$-models for small $n$. Chaotic
cases are denoted with \copyright. The billiard group of the $b_3=0$ compactification of the $B_2^{++}$-theory is the Weyl group of the hyperbolic Kac-Moody algebra $K_4$, whose Dynkin diagram is displayed in Table {\bf 13} (this algebra is denoted by ``$4$-$4$'' in \cite{SophieThesis}). Since $K_4$ is hyperbolic, the dynamics is chaotic. The billiard group $\mf{B}:=L_6$ which controls the $b_3=0$ compactification of the $B_4^{++}$ theory is a quotient of the formal Coxeter group $\mf{C}:= L_6^{\prime}$, whose Coxeter graph is displayed in Table {\bf 13}. The billiard region is not acute-angled and hence is not a fundamental domain of $L_6$. Moreover, there exist spacelike coweights and therefore the dynamics is non-chaotic. For all the non-chaotic compactifications in the table which are described by well known hyperbolic Weyl groups, the $\be$-space $\mf{M}_{\be}$ contains extra dilatonic directions which render the volume of the billiard region infinite.  }
\label{table:BnSmalln}
\end{center}
\end{table}

\begin{table}
\begin{center}
\begin{tabular}{ |m{25mm}|m{18mm}|m{18mm}|m{18mm}|}
\hline
$\{b_i(\mc{M})=0\}$ & $n=2$ & $n=3$ & $n=4$ \\
\hline\hline
$\{\},\{b_1\}$ & $A_3 \times A_1^+$ & $A_3^{+++}$ \copyright & $D_4^{++}$\copyright \\
$\{b_2\}$  & $A_1^{++}\times A_1$ & $M_5$  & $A_3^{++}$\\
$\{b_1,b_2\},\{b_2,b_3\},$ &  $A_1^{++}\times A_1$  & $A_2^{++}$ &$A_3^{++}$ \\
$\{b_1,b_2,b_3\}$ & $A_1^{++}\times A_1$  & $A_2^{++}$ &$A_3^{++}$ \\
$\{b_3\},\{b_1,b_3\}$ & $A_1^{+++}$ & $N_5$ &$A_3^{++}$  \\
\hline
\end{tabular}
\caption{Compactifications of the $D_n^{++}$-models for small $n$. Chaotic cases are denoted with \copyright. The $b_2=0$ compactification of the $D_3^{++}$ theory is governed by the billiard group $\mf{B}:=M_5$ which is a quotient of the formal Coxeter group $\mf{C}:= M_5^{\prime}$, as defined through the Coxeter graph in Table {\bf 13}. The billiard region contains obtuse angles and so is not a fundamental domain of $M_5$. Furthermore, the billiard region contains spacelike coweights and hence the dynamics is non-chaotic. The story for the billiard group $N_5$ is similar, and the Coxeter graph associated with the formal Coxeter group $N_5^{\prime}$ is given in Table {\bf 13}. The $b_1=0$ compactification of the $D_3^{++}$-theory is controlled by the Weyl group of the Lorentzian Kac-Moody algebra $A_3^{+++}$. Although this algebra is not hyperbolic, the theory is nevertheless chaotic since the number of dominant walls, $6$, exceeds the dimension of $\be$-space, dim$\ \mf{M}_{\be}=5$, and the billiard region is of finite volume. This is similar to the $X_9$ case in Table {\bf 1}. For all the non-chaotic compactifications in the table which are described by well known hyperbolic Weyl groups, the $\be$-space $\mf{M}_{\be}$ contains extra dilatonic directions which render the volume of the billiard region infinite.   }
\label{table:DnSmalln}
\end{center}
\end{table}

\begin{table}
\begin{center}
\begin{tabular}{|m{10mm}|m{40mm}|m{10mm}|m{40mm}|}
\hline
& & & \\
$K_4$ &  \includegraphics[width=30mm]{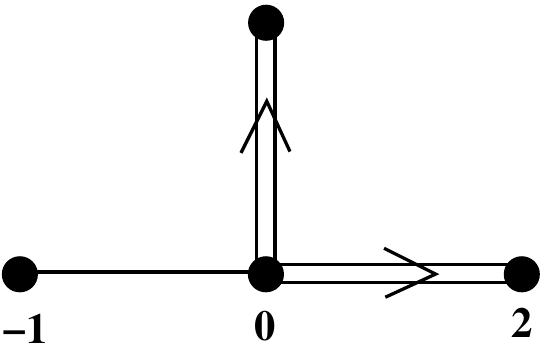} & $L_6^{\prime}$ &  \includegraphics[width=37mm]{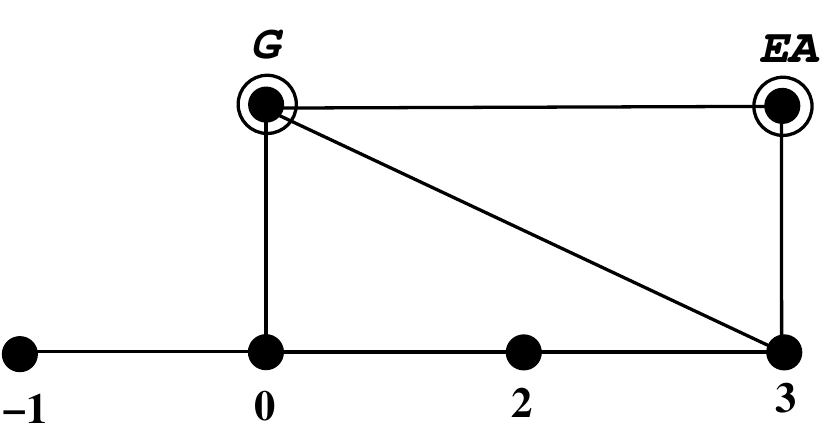} \\
& & & \\
 \hline
 & & & \\
$M_5^{\prime}$ &   \includegraphics[width=30mm]{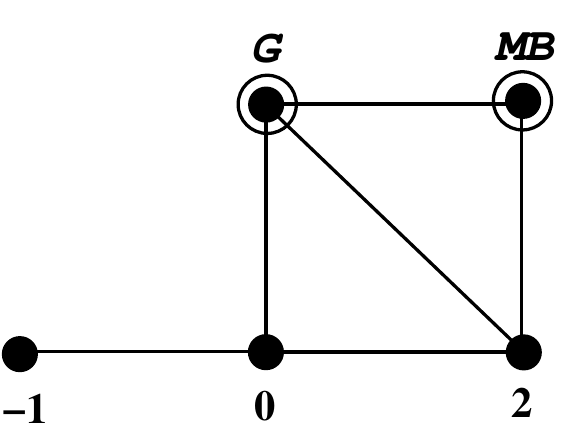} & $N_5^{\prime}$ & \includegraphics[width=30mm]{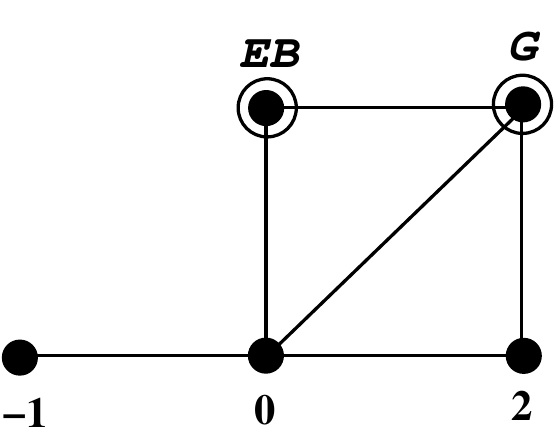} \\
& & & \\
\hline
\end{tabular}
\caption{$K_4$ represents the Dynkin diagram of a hyperbolic Kac-Moody algebra, whose Weyl group  controls the chaotic dynamics of the $b_3=0$ compactification of the $B_2^{++}$-theory. The remaining Coxeter graphs represent the formal Coxeter groups $L_6^{\prime}$, $M_5^{\prime}$ and $N_5^{\prime}$ associated with various compactifications of the $B_n^{++}$ and $D_n^{++}$ theories (see Tables {\bf 11} and {\bf 12}). As usual, the extra circles denote which dihedral angles are obtuse. In these graphs, ``$G$''Ê denotes the dominant gravity wall, ``$EA$'' the dominant electric wall of the Maxwell field $A$, ``$EB$'' and ``$MB$'' denote, respectively, the dominant electric and magnetic walls of the $2$-form $B$. See Tables {\bf 8} and {\bf 10} for a summary of the field contents of these theories. }
\label{table:Funny}
\end{center}
\end{table}

\end{appendix}

\end{document}